# New Research Trends in Electrically Tunable 2D van der Waals Magnetic Materials

Manh-Huong Phan

The Laboratory of Advanced Materials and Sensors, Department of Physics,

University of South Florida, Tampa, Florida 33620, USA

**The recent discovery of two-dimensional (2D) van der Waals (vdW) magnetic materials has provided new, unprecedented opportunities for both fundamental science and technological applications. Unlike three-dimensional (3D) magnetic systems, the electric manipulation of vdW magnetism (e.g., magnetization state, magnetic anisotropy, magnetic ordering temperature) down to the monolayer limit at ambient conditions enables high efficiency operation and low energy consumption, which has the potential to revolutionize the fields of spintronics, spin-caloritronics, and valleytronics. This article provides an in-depth analysis of the recent progress, emerging opportunities, and technical challenges in the electric manipulation of magnetic functionalities of a wide variety of 2D vdW magnetic systems ranging from metals to semiconductors and heterostructures. The state-of-the-art understanding of the mechanisms behind the electric modulation of magnetism in these 2D vdW magnetic systems will drive future research towards novel applications in spintronics, spin-caloritronics, valleytronics, and quantum computation.**

*Corresponding author: phanm@usf.edu (M.H.P)

# Table of Contents

## 1. Introduction

Magnetic materials are key components in spin-based electronic devices. Their applications are primarily driven by how their magnetic functionalities are tuned in response to external stimuli such as magnetic field, electrical field, strain, and light [1-6]. Instead of utilizing magnetic fields, electric fields can be used to manipulate magnetic functionalities (e.g., magnetization, magnetic anisotropy, and magnetic ordering temperature) in a ferromagnetic (FM) material in spintronic devices. For example, field-effect transistors can yield higher efficiencies and have lower energy consumption than their magnetically driven counterparts [7-10]. Electric field-modulated magnetic properties have been explored in a wide range of magnetic systems including FM metals [11,12], FM semiconductors [13], multiferroics [14,15], and magnetoelectric (ME) materials [16]. Since Ohno *et al.* demonstrated the electric field control of ferromagnetism in the semiconducting thin film (In,Mn)As in 2000 [13] and the large ME effect discovered in the multiferroic $BiFeO_3$ heterostructure by Ramesh *et al.* in 2003 [16], research on electrically controlled magnetism for spintronics applications has been rapidly advancing. Figure 1 illustrates this research trend.

Depending on the type of magnetic material, the mechanism of electrically controlled magnetism can be rather different [8-10,13-15]. For example, applying an electric field to a FM semiconductor such as (In,Mn)As [13] changes its carrier density, which in turn alters the magnetic interactions. As a result, the magnetic ordering temperature, magnetization, and magnetic anisotropy of the material can be tuned electrically. In a FM metal (e.g., Fe), however, the application of an electric field to the material through an insulating layer (e.g., MgO) will shift the Fermi level position at the metal/insulator interface (Fe/MgO), altering the magnetic anisotropy of the FM metal (Fe) [11,17]. In a multiferroic (e.g., $BiFeO_3$) in which both magnetism and electricity are coupled, the magnetization can be manipulated by electric fields, and vice versa the electric polarization can be manipulated by magnetic fields [2,7,15,16]. The majority of research in this field is aimed at creating new materials whose magnetic functionalities can be manipulated or triggered at ambient temperature by small electric fields, enabling the development of ultralow-power and high-performance spintronic devices [18,19]. Since downsized spintronic devices are in high demand, it is essential to reduce the dimensions of magnetic materials used in these devices [10,18-33]. In most cases, however, the magnetic properties of the materials are drastically degraded upon size reduction to the nanoscale, posing one of the most challenging issues in the field.

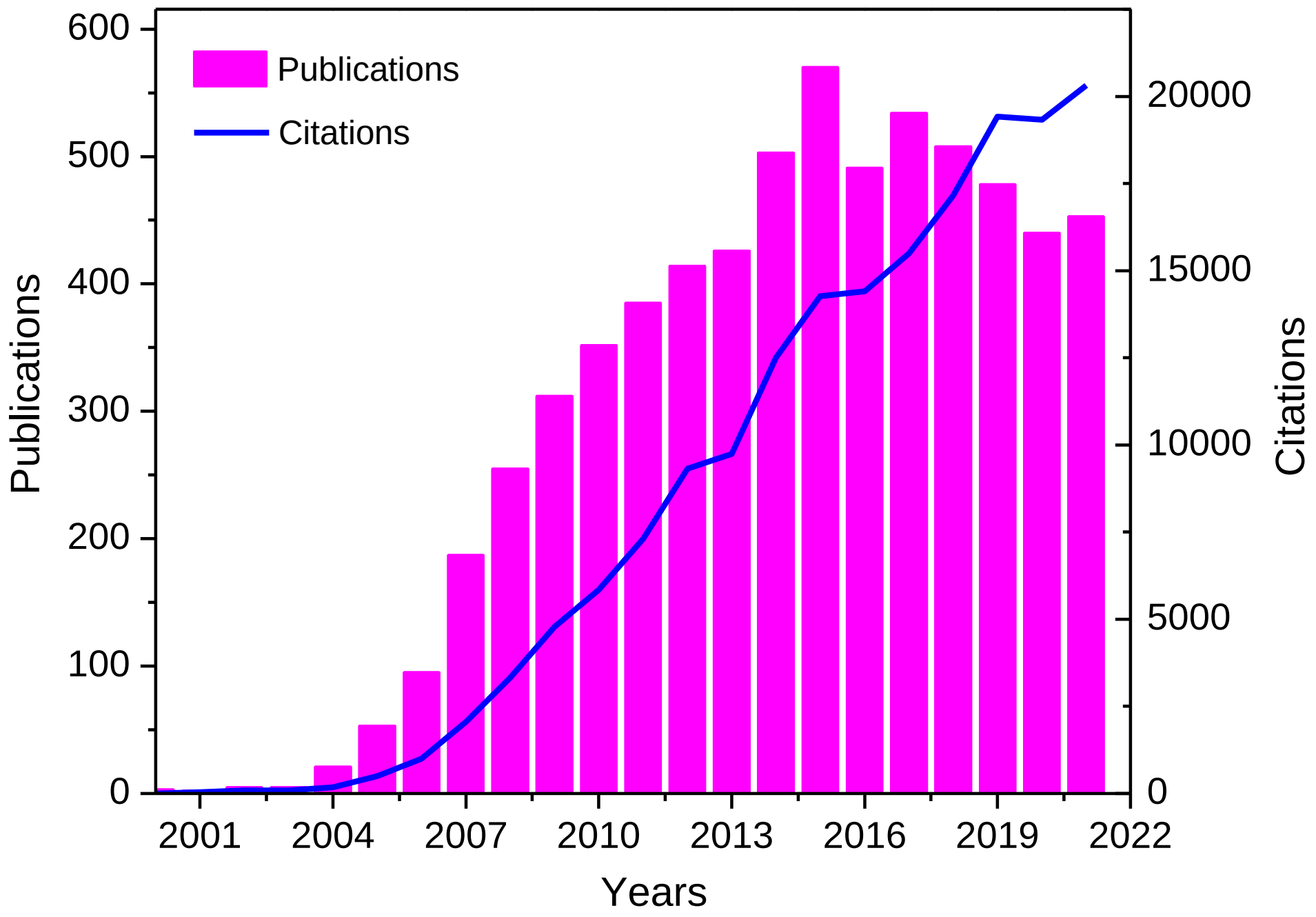

**Figure 1.** Number of publications and citations per year in the research of electrically tunable magnetic materials. The data were collected from Web of Science with “electric control of magnetism” or “multiferroic” as a keyword.

The recent discovery of two-dimensional (2D) van der Waals (vdW) magnetic materials has enabled the electric manipulation of magnetism down to the monolayer limit, which may pave the way towards resolving the "downsizing" and related issues [10,20-33]. The long-range magnetic ordering reported in $CrI_3$ monolayers of mechanically exfoliated crystals in 2016 [36] acted as a catalyst in the field of 2D vdW magnetism, which has been rapidly advancing ever since. As a result, a large pool of new 2D vdW magnetic materials such as $Cr_2Ge_2Te_2$ [37], $Fe_3GeTe_2$ [38], $Fe_5GeTe_2$ [39], $VSe_2$ [40], $V_5Se_8$ [41], $MnSe_2$ [42], $CrBr_3$ [43], $CrCl_3$ [44], $CrSe_2$ [45], $CrTe_2$ [46], $Cr_2Te_3$ [47], $Cr_3Te_4$ [48], and $TMPS_3$ ($FePS_3$, $MnPS_3$, $NiPS_3$, $CoPS_3$, $FePSe_3$, $MnPSe_3$) [10,28,49] have been discovered. This collection forms a diverse family of low-dimensional magnetic materials with new and novel functionalities for ultra-low power and ultra-compact spintronics, valleytronics, and quantum computing devices. It is worth noticing that most of the practical device applications require magnetic functionality at ambient conditions. While the magnetic properties of 2D $CrI_3$ and $Cr_2Ge_2Te_2$ have been extensively studied during the last few years [10,28,36,37], these 2D vdW magnets are restricted to operate at cryogenic temperatures (below 100 K). In this context, 2D vdW magnets with much higher Curie temperatures, near room temperature (e.g., $Fe_3GeTe_2$ [38], $CrTe_2$ [46]) or above room temperature (e.g., $VSe_2$ [40], $MnSe_2$ [42], $Fe_5GeTe_2$ [39]), are of practical importance. The most recent discovery of tunable room temperature ferromagnetism in 2D dilute magnetic semiconductors (2D-DMSs) based on semiconducting transition metal dichalcogenides (TMDs) doped with magnetic atoms, such as V-$WS_2$ [50], V-$WSe_2$ [51,52], and Fe-$MoS_2$ [53] monolayers, has attracted a great deal of attention. Interestingly, these newly discovered 2D-DMSs exhibit strong and tunable magnetic responses to light irradiation [54] and electric gating [52,55]. While a full understanding of the physical mechanisms that govern the electric manipulation of magnetism in 2D vdW magnets and their heterostructures has remained largely elusive, this knowledge is the key to unlocking the doors to 2D vdW spintronics, spincaloritronics, and valleytronics (Fig. 2). Despite a number of excellent reviews that have focused mostly on the outstanding magnetic, magneto-optic, and spin transport properties of 2D vdW magnets and heterostructures [10,18-35], a critical, in-depth analysis of their electrically tunable magnetic properties and associated phenomena is lacking.

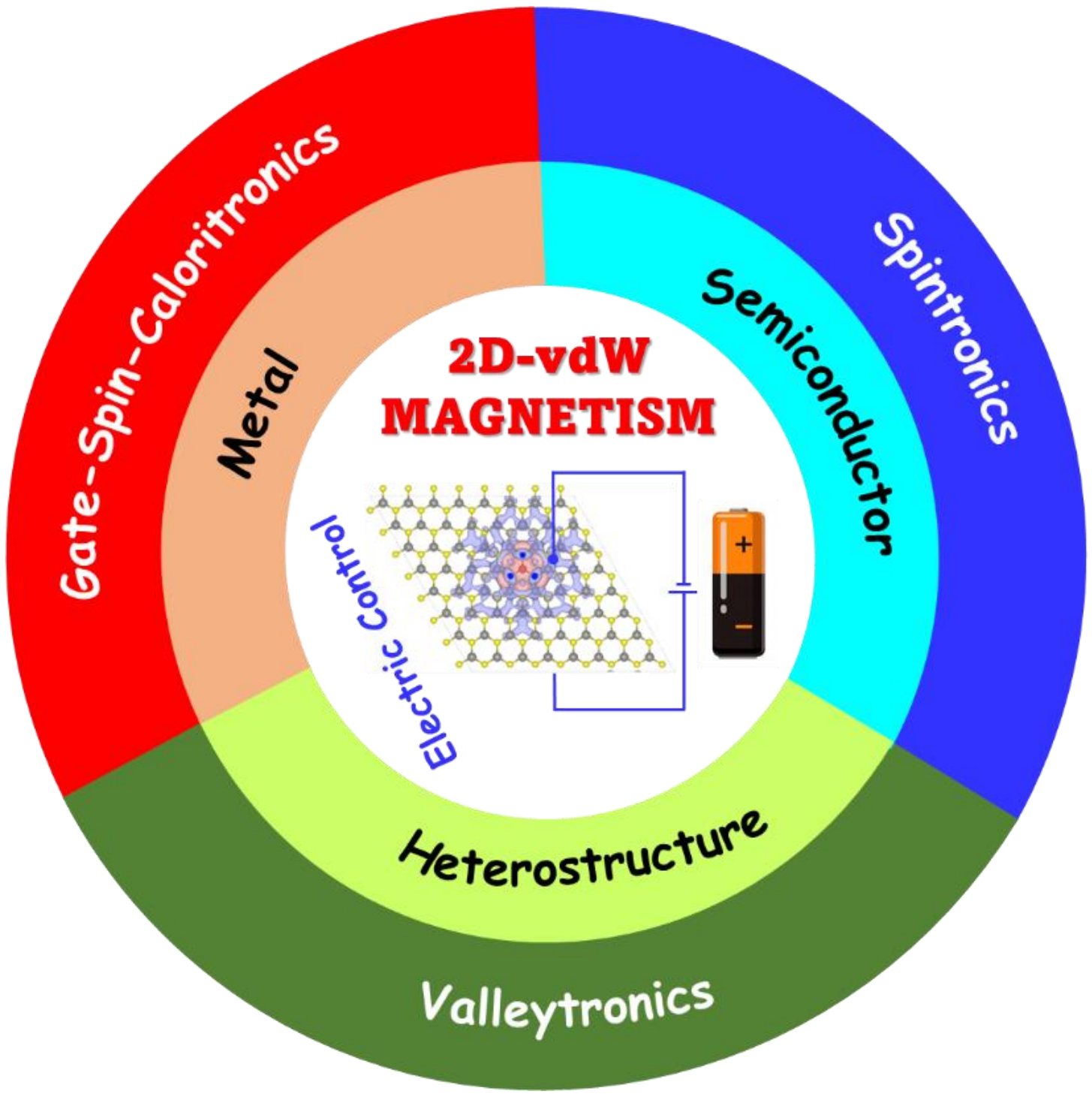

**Figure 2.** Potential applications of electrically controllable two-dimensional van der Waals magnets and heterostructures for spintronics, spin-caloritronics, and valleytronics.

The overall aim of this review is to provide the state-of-the-art understanding of the mechanisms of electric manipulation of magnetism in the experimentally and theoretically discovered 2D vdW magnets and their heterostructures through a comprehensive analysis of the most recent studies on the electric (electric gating, unless specified) manipulation of magnetic functionalities in these systems. After a brief description of types of magnetic interactions and fundamentals of electrically-modulated magnetism, the electric field-dependent magnetic properties of 2D vdW magnetic metals, semiconductors, and their heterostructures are systematically assessed. Emerging opportunities and technical challenges are discussed. The article concludes with an outlook on the future research in this new but rapidly expanding research field.

## 2. Electric manipulation of two-dimensional vdW magnetism

### 2.1. Fundamentals

Magnetic ordering in a magnetic material is determined by exchange interactions between magnetic moments or spins that are coherently arranged in a material. The material can exhibit a ferromagnetic (FM) order, antiferromagnetic (AFM) order, ferrimagnetic (FiM) order, or spin textures like helimagnetism and skyrmions, or some combination (Fig. 3a).

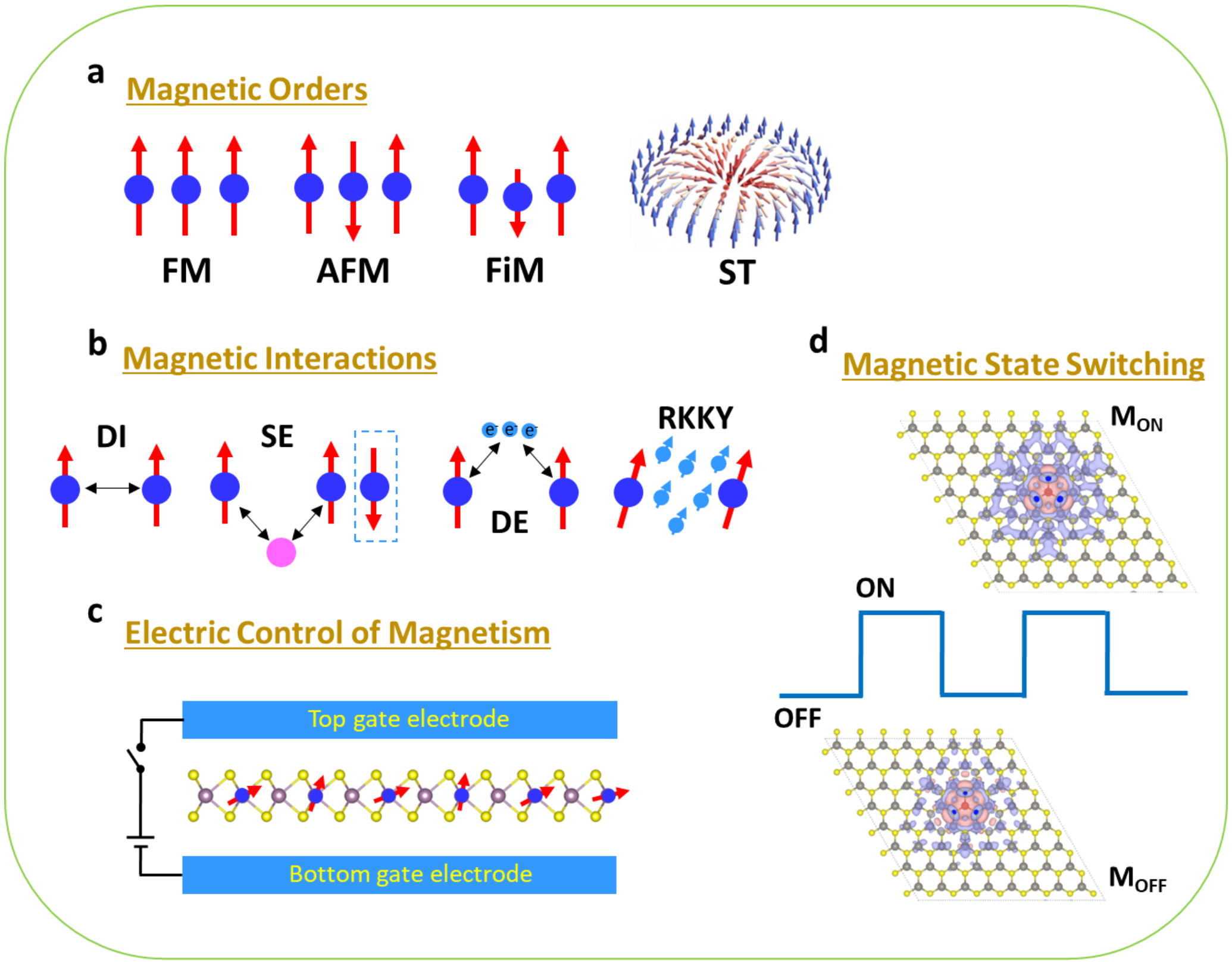

**Figure 3.** (**a**) Different types of magnetic ordering including ferromagnetic (FM), antiferromagnetic (AFM), ferrimagnetic (FiM), and spin texture (ST) orderings. (**b**) Different

mechanisms of magnetic interactions including direct interaction (DI), super-exchange (SE), and double exchange (DE) interactions. For SE interaction, either FM or AFM coupling can exist. (**c**) Illustration of the electric gating of magnetization. (**d**) Application of an electric field switches the weak FM state (OFF, low magnetization) into the strong FM state (ON, high magnetization), for example, on V-doped TMD monolayers (e.g., V-$WS_2$, V-$WSe_2$). Depending on the magnetic nature of a material, the application of an electric field to the material can change PM to FM, or AFM to FM, or FiM to FM, or FM1 to FM2.

Although the Mermin-Wagner theorem predicts the absence of long-range magnetic order in 1D or 2D isotropic magnetic systems [56], the magnetic ordering can be established if a significant magnetic anisotropy (e.g., magnetocrystalline anisotropy) is introduced into the material as it opens a gap for magnon excitations that suppress thermal fluctuations [10,28]. It has been shown that although the magnitude of this anisotropy is small, it still has a crucial influence on magnetic ordering in 2D magnetic systems [36-38] where Mermin-Wagner conditions are unfulfilled. In addition to magnetocrystalline, surface, and interface anisotropies, local magnetic anisotropy that arises from various sources of defects, interfaces, and surfaces can have considerable impact on magnetic ordering in these low-dimensional magnetic systems [10,57]. Recently, Santos *et al.* showed that in finite-size 2D vdW magnetic systems (within millimeters), short-range magnetic interactions can be sufficient to establish a stable magnetic ordering state at finite temperatures [58]. This provides an explanation for the occurrence of magnetic ordering in several 2D vdW magnetic systems with small magnetic anisotropies, including magnetic 2D-TMDs [50-53,57]. In general, the exchange interaction between spins in a simple magnetic system can be described by the Heisenberg spin Hamiltonian,

$$H_{ex} = -\sum_{i,j}^{n} J_{i,j} S_i S_j, \qquad (1)$$

where $J_{i,j}$ is the exchange constant that depends on the separation between two spins and its sign determines types of magnetic interaction ($J_{i,j} > 0$ for FM interaction, and $J_{i,j} < 0$ for AFM interaction). If other terms such as crystal field are added to the Hamiltonian, complex magnetic order such as helimagnetism or skyrmions can occur [59,60]. Four types of magnetic interactions are often considered in 2D magnetic systems (Fig. 3b) including *direct* interaction (DI), *super-exchange* (SE) and *double-exchange* (DE), and Ruderman-Kittel-Kasuya-Yosida (RKKY) interactions. Extended SE, Super-SE, and multi-intermediate DE interactions are also considered in complex magnetic systems [10,28]. DI is generally known as the main mechanism for ferromagnetism in magnetic metals such as Fe and $Fe_3GeTe_2$ [38] in which conduction electrons are fully delocalized and hybridized with magnetic ions. The DI originates from the overlap of electronic wave functions of two neighboring magnetic atoms. Owing to the strong DI between individual spins, metallic vdW magnets such as $Fe_3GeTe_2$ [38] often exhibit higher magnetic ordering temperatures (Curie temperature, $T_C$) as compared to insulating or low-conducting vdW magnets like $CrI_3$ [36] or $Cr_2Ge_2Te_6$ [37]. In cases of metals with localized magnetic moments (e.g., Gd) where the mediators are conduction electrons, a localized magnetic moment can induce spin polarization to its surrounding conduction electrons and this polarization couples to the next neighboring localized moment. This occurs via an indirect exchange process, following the RKKY interaction (Fig. 3b). On the other hand, SE interaction can be either ferromagnetic or antiferromagnetic, and the sign and strength of SE coupling varies depending on several factors (e.g., bonding angle and distance). The SE theory is well suited to describe isolated spins that are present in magnetic insulators such as $LaMnO_3$ [61] and $Y_3Fe_5O_{12}$ (YIG) [62] as well as 2D low-conducting semiconductors like $CrI_3$ [36] and $Cr_2Ge_2Te_6$ [37]. These isolated spins interact with each other via non-magnetic ions. For instance, the magnetic moments of $Cr^{3+}$ ions in $CrI_3$ are coupled ferromagnetically via I ions by the SE mechanism [36]. In another interesting case, $LaMnO_3$, an antiferromagnetic insulator, has electrons possessing antiparallel spins in the two opposing lobes of the $O^{2-}$ p-orbital and so the energy savings provided by orbital hybridization translates to an effective AFM coupling between the coordinating $Mn^{3+}$ ions on either side of the oxygen ion [61]. However, when $La^{3+}$ ions are substituted by $Sr^{+}$ ions in $La_{1-x}Sr_xMnO_3$, the total charge must be reduced to retain a neutral charge condition thus converting $Mn^{3+}$ ions partially into $Mn^{4+}$ ions. The transfer of an itinerant $e_g$ electron between neighboring $Mn^{3+}$ and $Mn^{4+}$ sites (local $t_{2g}$ spins) through the $O^{2-}$ anion results in a FM interaction due to on-site Hund's rule coupling. This process occurs simultaneously, which is coined as the DE interaction. Both SE and DE theories have been widely applied to interpret the magnetism and conductivity in many complex magnetic oxide systems like doped manganites [61,63] and cobaltites [64] $Ln_{1-x}R_xBO_3$ (*Ln* = La, Nd, Pr; *R* = Ca, Sr, Ba; *B* = Mn, Co).

Based on density functional theory (DFT) calculations including the Hubbard *U*-term, Shu *et al.* showed that the magnetic exchange coupling of Fe moments in Fe-doped $MoS_2$ nanosheets depended on the competition between DE and SE interactions [65]. In the case of an Fe-doped $MoS_2$ monolayer, the magnetic moments of substitutional Fe dopants at Mo sites are ferromagnetically coupled via the DE mechanism. For Fe-doped $MoS_2$ bilayers or multilayers, however, the *d*-electron interaction among Fe dopants favors AFM coupling via the SE mechanism, possibly due to the formation of intercalated and substitutional Fe complexes in these structures. Nonetheless, Zhang *et al.* demonstrated theoretically and experimentally for V-doped $TX_2$ (*T* = W, *X* = S, Se) monolayers that in addition to the dominant FM couplings between V-moments, AFM coupling emerged for V moments at close V-V distances due to orbital hybridization, arising from V-dopant aggregations [50]. In such magnetically-doped TMD monolayers, long-range FM interaction between V spins is also suggested to be mediated by itinerant spin-polarized holes via the RKKY mechanism [52,66]. It is likely that the exchange interaction among magnetic dopants is mediated by defect-polarized spins [67], hence, the DE, SE, or RKKY model alone cannot account for the observed 2D-TMD magnetism.

In addition to the DI, DE, SE, and RKKY interactions, antisymmetric exchange, also known as the Dzyaloshinskii-Moriya interaction (DMI), can also contribute to the total exchange interaction between two neighboring magnetic spins and drive the formation of topological spin textures such as magnetic skyrmions (Fig. 3a) in a wide range of magnetic systems [59,60]. Magnetic skyrmions, which are known as statically stable solitons experimentally observed in bulk MnSi [68] or magnetic thin films [69], are an emerging platform for ultrahigh-density spin memory device applications [59,60]. In some 2D vdW magnets such as $Fe_3GeTe_2$ and related systems, strong spin-orbit coupling (SOC) has been reported to retain the perpendicular magnetic anisotropy down to the monolayer limit as well as to lift chiral degeneracy, leading to the formation of magnetic skyrmions through the DMI mechanism [70-79]. Indeed, Néel-type skyrmions have been observed in $Fe_3GeTe_2$ when interfaced with another material such as $WTe_2/Fe_3GeTe_2$ [70] and $hBN/Fe_3GeTe_2$ [71]. Recently, a novel van der Waals interface composed of $Fe_3GeTe_2$ and $Cr_2Ge_2Te_6$ has been reported to host two different groups of magnetic skyrmions [79]. In this case, the inversion symmetry breaking near the $Fe_3GeTe_2/Cr_2Ge_2Te_6$ interface was believed to give rise to the DMI, creating the two distinct magnetic skyrmions on both sides of the individual components. Electric and magnetic control of such skyrmions is of potential interest for spintronics and computing applications [80].

Since the magnetic exchange interactions in a magnetic material can be altered by varying distances between magnetic ions and/or carrier concentration, it is possible to employ electric fields to manipulate carrier (electrons or holes) concentration (in case of FM semiconductors) or shift Fermi levels (in case of FM metals) to change the overall magnetic response of the material [9,10,23]. To realize this, a typical field-effect transistor-type device utilizing such a magnetic material can be designed and fabricated (Fig. 3c). It has been reported that the application of electric fields can switch a magnetic state from PM to FM, AFM to FM, FiM to FM, or FM1 to FM2 (Fig. 3d). One typical example of this is the magnetic semiconductor (Ga,Mn)As [9] of which both the magnetization and Curie temperature can be effectively manipulated by electric fields. In this case, the applied electric field changes the concentration of holes, which, in turn, mediate the *p*-*d* exchange interaction between the Mn moments and holes, thus affecting the magnetic order and the net magnetization of the material. In a similar fashion, the magnetization of a V-doped $WSe_2$ monolayer–a new type of 2D diluted magnetic semiconductor–has recently been manipulated by electric gating [52,55]. Considerable changes in magnetization, magnetic anisotropy, and magnetic ordering temperature have been observed in several magnetic vdW systems and heterostructures [10,27-29], and the phenomena of which are discussed in detail below.

## 2.2. Two-dimensional vdW magnetic materials

### *2.2.1. Metals*

Recall that an application of an electric field to a magnetic metal through an insulating layer can shift its Fermi level at the metal/insulator interface, which changes the occupancy of orbitals and hence its effective magnetic anisotropy. This effect was reported in Fe/MgO and Fe(Co)/MgO systems [11,17]. A similar effect has recently been observed in vdW magnetic metals [81-83].

One of the most widely studied metallic vdW magnetic systems is the itinerant ferromagnet $Fe_3GeTe_2$ because of its high Curie temperature ($T_C$ ~200-230 K for the bulk and ~130 K for the monolayer [38]) and strong magnetic responses to external stimuli (electric fields, strain, light) [81,82,84,85]. $Fe_3GeTe_2$ crystalizes in the P63/mmc space group with one $Fe_3Ge$ layer sandwiched by two Te layers (Fig. 4a) [38]. The separation between two adjacent monolayers is ~2.95 Å. The valence states of $Fe_3GeTe_2$ can be viewed as $(Te^{2-})(Fe_I^{3+})[(Fe_{II}^{2+})(Ge^{4-})](Fe_I^{3+})(Te^{2-})$ with inequivalent $Fe_I^{3+}$ and $Fe_{II}^{2+}$ sites within the $Fe_3Ge$ plane. The ferromagnetism of the material is derived from two inequivalent $Fe_I^{3+}$ and $Fe_{II}^{2+}$ sites within the monolayer.

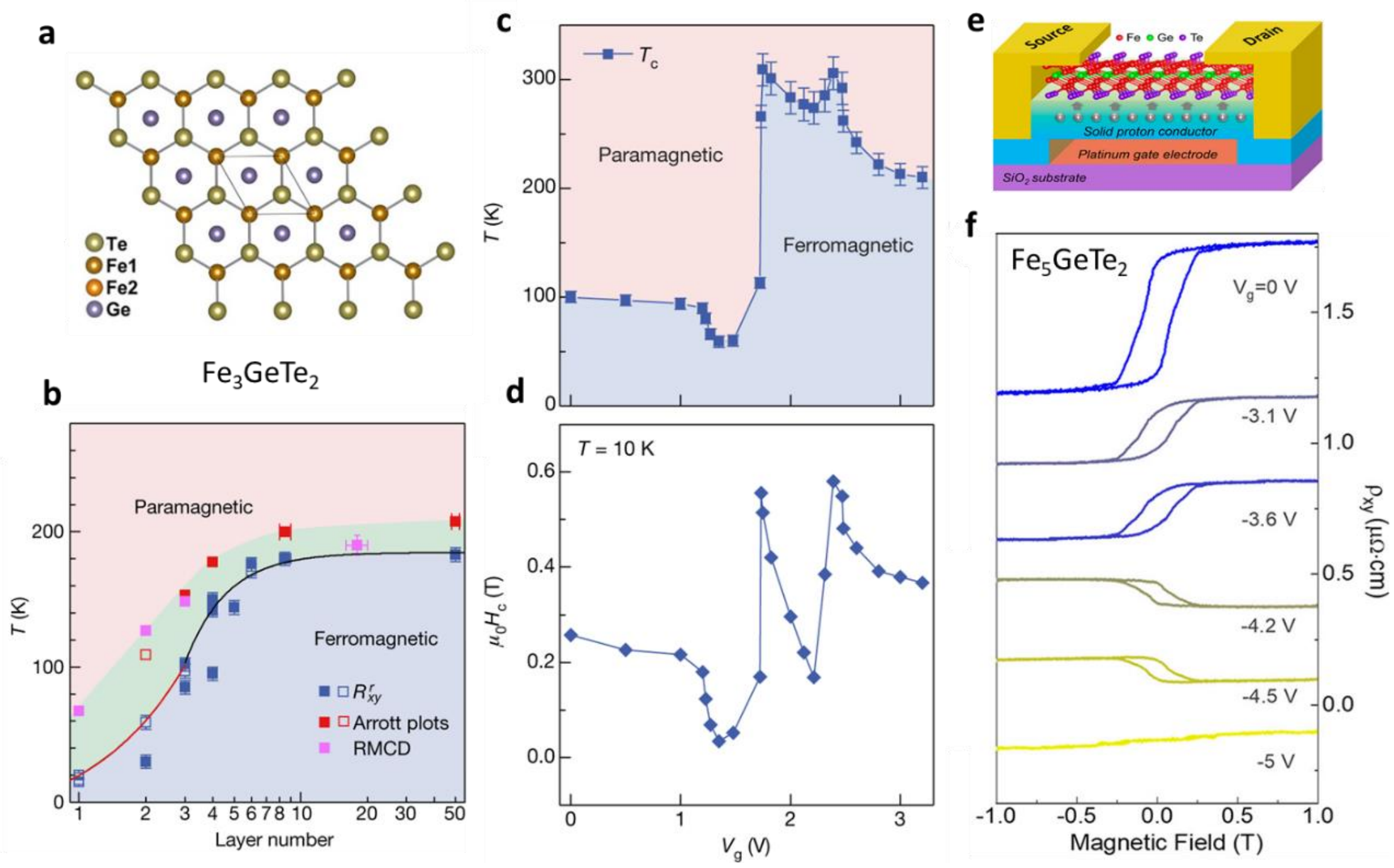

**Figure 4.** (**a**) Top view of the $Fe_3GeTe_2$ structure. (**b**) Change in the magnetic ordering temperature ($T_C$) of $Fe_3GeTe_2$ with layer number; panels (a,b) reproduced with permission from Ref. [38]. Changes in (**c**) Curie temperature ($T_C$) and (**d**) coercive field ($H_C$) of a trilayer $Fe_3GeTe_2$ upon ionic gating; panels (c,d) reproduced with permission from Ref. [81]. (**e**) A protonic gating device for $Fe_5GeTe_2$, and (**f**) changes in the magnetic loops $\rho_{xy}(H)$ upon protonic gating. Panels (e,f) reproduced with permission from Ref. [83].

In bulk $Fe_3GeTe_2$, however, partially filled *d*-orbitals of Fe atoms dominate the energy band structure around the Fermi level, governing the electronic itinerancy and local FM. It has been reported that bulk $Fe_3GeTe_2$ exhibits a large out-of-plane magnetic anisotropy with respect to the vdW planes, and the strength of this anisotropy remains significant in the monolayer, thus preserving long-range FM order [38]. The Curie temperature of $Fe_3GeTe_2$ varies significantly with thickness when the number of layers is less than ~10 but becomes almost unchanged for thicker films (Fig. 4b). When the thickness of $Fe_3GeTe_2$ is less than 15 nm, a single PM to FM transition is observed. For thicker films, however, a distinct magnetic behavior emerges in the intermediate temperature range (130-207 K), which has been attributed to the formation of labyrinthine domain patterns [38].

DFT calculations have predicted that applying an electric field to $Fe_3GeTe_2$ causes the Fermi level to rise to an area with a large density of states (DOSs), affecting the exchange interaction and the magnetic anisotropy [86]. A detailed analysis of the electronic structure of $Fe_3GeTe_2$ highlights the importance of the occupation and splitting of the Te($p_z$)-Fe($d_{z2}$) bond states in modifying the magnetic anisotropy. $Fe_3GeTe_2$ possesses a very high electron density of ~$10^{14}$ $cm^{-2}$; however, it is experimentally impractical to apply substantial electric fields to the material to alter the charge density and hence its magnetic functionality. This large electric field barrier occurs primarily due to field screening effects by the charge carriers in the metal. To overcome this, Deng *et al.* developed an $Al_2O_3$-assisted exfoliation method that enables the isolation of $Fe_3GeTe_2$ monolayers from its bulk crystal and the realization of gate voltage-controlled magnetism in these 2D samples [81]. Using an ionic field-effect transistor with a solid electrolyte ($LiClO_4$ dissolved in polyethylene oxide matrix), the authors demonstrated that the applied positive gate voltage could significantly modulate the ferromagnetism in the $Fe_3GeTe_2$ nanoflakes (four layers) (Fig. 4c,d). Interestingly, the application of $V_g$ ~ 2 V increased the $T_C$ of the material to 300 K (Fig. 4c) making it applicable for gate voltage-controlled vdW spintronics under ambient conditions. The gate voltage dependence of coercivity (Fig. 4d) was also observed to follow that of $T_C$ (Fig. 4c). The large changes in $T_C$ and $H_C$ caused by positive gate voltages have been attributed to the large changes in the DOS at the Fermi level. The non-monotonic dependence of $T_C$ on gate voltage (> 2V) seems to suggest the Fermi level passes by the peak point. The Stoner model was employed to interpret the mechanism of the ionically gated magnetism in the 2D $Fe_3GeTe_2$. When using lithium-ion conducting glass-ceramics (LICGC) as a solid $Li^+$ electrolyte in an ionic gating device, however, Chen *et al.* observed weakened ferromagnetism in the $Fe_3GeTe_2$ nanoflakes [87]. When applying a gate voltage, $Li^+$ ions were introduced into $Fe_3GeTe_2$, causing considerable decreases in both $H_C$ and $T_C$. Indeed, the application of a positive gate voltage $V_g$ = 3.5 V at $T$ = 100 K was found to decrease $H_C$ by ~ 25% from the reference voltage ($V_g$ = 0 V). Such noticeable variations in $H_C$ and $T_C$ are attributed to intercalations of $Li^+$ ions between the atomic layers of $Fe_3GeTe_2$.

Zheng *et al.* also reported modulation in the interlayer magnetic coupling of $Fe_3GeTe_2$ nanoflakes upon the application of protonic gates [88]. The increase of protons intercalated among vdW layers was found to enhance the interlayer magnetic coupling, inducing a large zero-field cooled (ZFC) exchange bias (EB) effect ($H_{EB}$ ~ 1.2 kOe) in the protonated $Fe_3GeTe_2$ nanoflakes in which both FM and AFM interactions coexisted and competed. Also utilizing a solid protonic gate device (Fig. 4e), Tan *et al.* demonstrated that an extremely high electron doping concentration of ~$10^{21}$ $cm^{-3}$ can be achieved in $Fe_5GeTe_2$ nanosheets, and a magnetic phase transition from the FM to AFM state emerges due to the protonic gating effect (Fig. 4f) [83]. These findings demonstrate the possibility of enhancing interlayer magnetic couplings in vdW metallic magnets like $Fe_3GeTe_2$ and $Fe_5GeTe_2$ by proton intercalation, expanding the applications of these 2D vdW magnets.

In a recent work, Yang *et al.* proposed the design of a spin-valve device comprising monolayer $Fe_3GeTe_2$ and two ferromagnetic electrodes and investigated the spin-dependent transport characteristics of this device using *ab initio* quantum transport simulation [89]. They found a high magnetoresistance of ~390% and a significant increase of MR to 450-510% upon electrical gating. Experimental studies are needed to validate this prediction. The effect of current on the ferromagnetism of $Fe_3GeTe_2$ has also been reported [90,91]. Zhang *et al.* showed that an in-plane current could tune the magnetic state in a 2D $Fe_3GeTe_2$ film from a hard magnetic to a soft magnetic state [91]. In an $Fe_3GeTe_2$/Pt structure, the magnetization of the 2D $Fe_3GeTe_2$ layer could be effectively switched by spin-orbit torque (SOT) originating from current flowing in the Pt layer. These studies pave the way for developing a new generation of SOT-based spintronic devices [92]. One of the most challenging issues involved with all $Fe_3GeTe_2$-based devices is that this material is extremely air sensitive and its intrinsic magnetic property rapidly degrade or even disappear when exposed to air [93]. A recent review by Phan *et al.* pointed out the significant effect of oxidization on interfacial magnetic coupling and exchange bias in $Fe_3GeTe_2$ and its heterostructures [35].

Besides $Fe_3GeTe_2$, other metallic 2D vdW ferromagnets like $VSe_2$, $MnSe_2$, and $CrSe_2$ have been extensively studied over the last few years [40,42,45]. While bulk $VSe_2$ crystalizes in a 1T structure and shows paramagnetic characteristics down to low temperature [94], its 1T-monolayer counterpart has been reported to exhibit ferromagnetic ordering above room temperature (RT) [40,95]. Although there is an ongoing debate on the origin of the observed ferromagnetism in $VSe_2$ monolayers [40,95-104], recent studies have suggested both intrinsic (V spins) and extrinsic (Se vacancy-induced spins) sources of magnetism in these films [40,95]. While decoupling the intrinsic magnetism from the defect-induced (extrinsic) magnetism is a challenging task, the latter contributes more dominantly to the magnetism of this 2D system [95,104]. Although no experimental report was found to show the electric field-dependent magnetic property of monolayer $VSe_2$, Gong *et al.* predicted a strong magnetic response of bilayer $VSe_2$ to an external electric field [105]. It is worth noticing that bilayer $VSe_2$ was assumed to have a 2H structure and exhibit an antiferromagnetic order [105], while MBE-grown multilayer to single layer films of $VSe_2$ were experimentally reported to crystalize in a 1T structure and exhibit ferromagnetic order [40]. Therefore, it would be interesting to both theoretically and experimentally examine if the RT ferromagnetism of 1T $VSe_2$ monolayer or bilayer films can be tuned by electric fields.

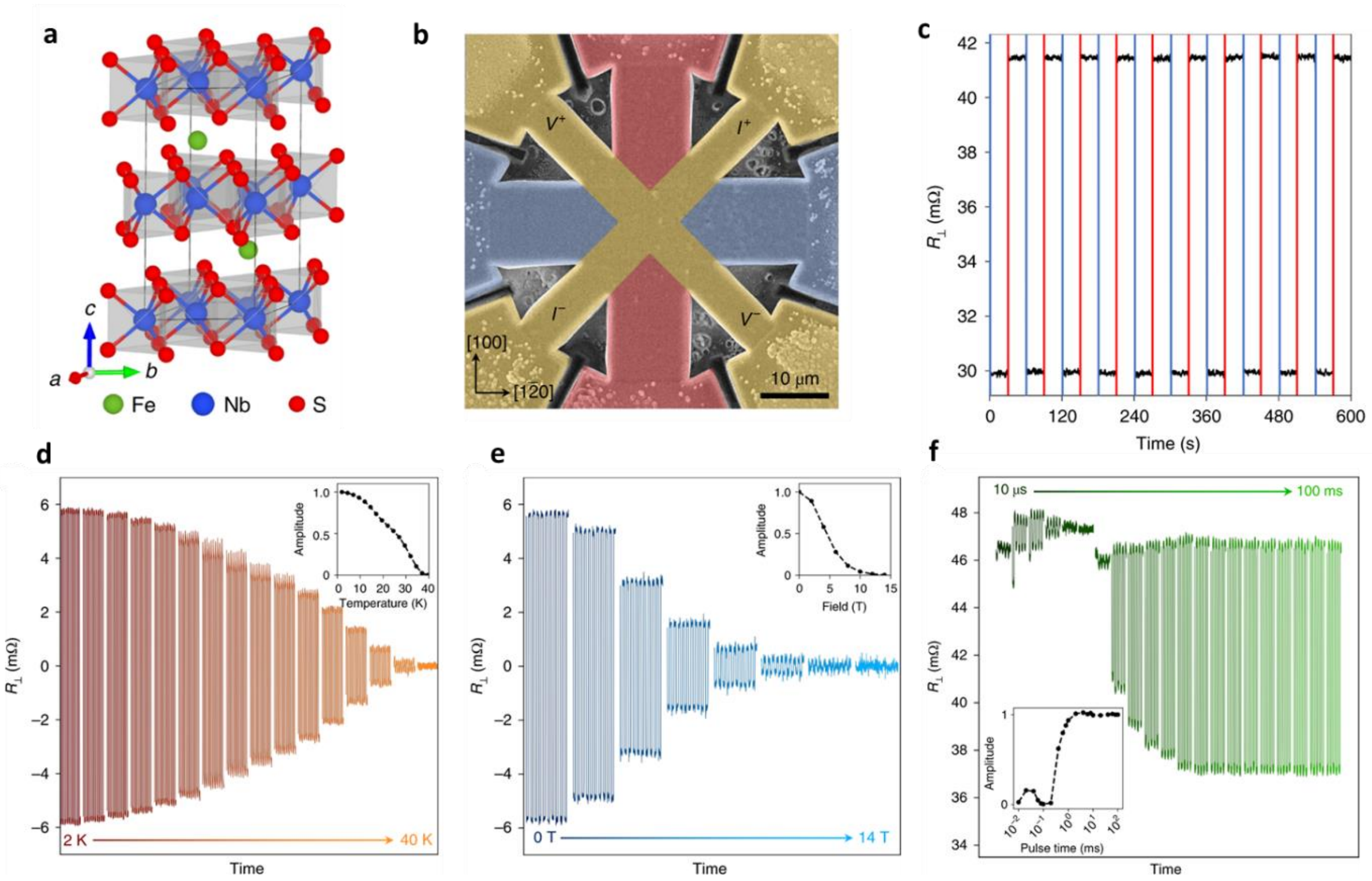

**Figure 5.** (**a**) The crystal structure of $Fe_{1/3}NbS_2$. (**b**) SEM image of the $Fe_{1/3}NbS_2$ switching device. (**c**) Resistance ($R_{\perp}$) switching between two states when orthogonal current pulses are applied. (**d**) The resistance switching behavior is suppressed by temperature and disappears above the Néel temperature (42 K). (**e**) The resistance switching behavior at 2 K is suppressed by magnetic fields.

(**f**) The time-dependent resistance switching behavior taken at 2 K under various current pules. Reproduced with permission from Ref. [108].

Unlike $Fe_3GeTe_2$ and $VSe_2$, $CrSe_2$ is an air-stable material and has been reported to preserve its ferromagnetic character over time [45]. $CrSe_2$ crystalizes in a 1T structure and exhibits ferromagnetic order below ~110 K. It has been reported that ferromagnetism weakens rapidly when the thickness of the $CrSe_2$ film is reduced to the monolayer, with a corresponding decrease of its Curie temperature from ~110 K for the bulk to ~65 K for the monolayer [45]. While bulk $CrSe_2$ exhibits a strong out-of-plane magnetic anisotropy, this anisotropy is considerably weakened in its monolayer counterpart. Theoretical calculations predict the metallic magnetism of monolayer $CrSe_2$ (also $FeSe_2$) with out-of-plane magnetic anisotropy, and the application of an external electric field can change the magnetic anisotropy from the out-of-plane to the in-plane direction [106]. However, gate-modulated magnetic experiments are needed to validate this prediction. Belonging to the class of $Cr_xX_y$ ($X$ = Se, Te), $CrTe_2$ and $Cr_2Te_3$ have also been reported to exhibit ferromagnetic ordering near room temperature [46,47] and therefore are perspective candidates for exploiting electrically tunable magnetoelectronics.

$Fe_{1/3}NbS_2$ emerges as a magnetically intercalated layered transition metal dichalcogenide that exhibits antiferromagnetic order ($T_N$ ~42 K), a metal-insulator/semiconductor transition around the $T_N$, exchange bias, and exotic spin-transport phenomena [107,108]. The magnetic ordering of 1/3 fractional intercalation of transition metal atoms between the layers of 2H-$NbS_2$ (Fig. 5a) can vary from antiferromagnetism (e.g., $Fe_{1/3}NbS_2$) to helimagnetism (e.g., $Cr_{1/3}NbS_2$). Unlike $Cr_{1/3}NbS_2$ [109], $Fe_{1/3}NbS_2$ orders antiferromagnetically below 42 K [108], with strong magnetic anisotropy along the $c$-axis. This TMD antiferromagnet becomes a promising candidate for use in spintronic devices with fast switching times, insensitivity to magnetic perturbations, and reduced crosstalk. Recently, Nair *et al.* demonstrated electric switching of magnetic order in single crystals of $Fe_{1/3}NbS_2$ (Fig. 5) [108]. The in-plane component of the AFM order in $Fe_{1/3}NbS_2$ can be rotated by current pulses via an anti-damping-like spin transfer torque. This enables coding of electronic information into the AFM state, which can then be read out with electronic resistivity measurements via zero-field AMR. These findings lay the foundation for the development of electronically controlled AFM memory devices. Figure 5 shows the structure of an $Fe_{1/3}NbS_2$-based switching device and its resistance-switching behavior where the sharp switching of the magnetic state is achieved and manipulated by electric means (Fig. 5b-f). However, the operation of this device is limited to temperatures below the Néel temperature of $Fe_{1/3}NbS_2$ ($T_N$ ~42 K) [108]. Therefore, there are research opportunities for TMD antiferromagnets with high Néel temperatures around ambient temperature.

#### 2.2.2. Semiconductors

##### *2.2.2.1. Low temperatures*

Unlike ferromagnetic metals, the electrical control of both charge and spin degrees of freedom in ferromagnetic semiconductors make them an ideal candidate for use in spintronic devices. The recent discovery of intrinsic magnetism in 2D vdW magnetic semiconductors such as $CrI_3$ [36], $CrBr_3$ [43], and $Cr_2Ge_2Te_6$ [37] has provided a new and exciting opportunities for exploring electrically tunable magnetism and magnetoelectric phenomena down to the 2D limit and towards the development of next generation 2D vdW spintronic and valleytronic devices.

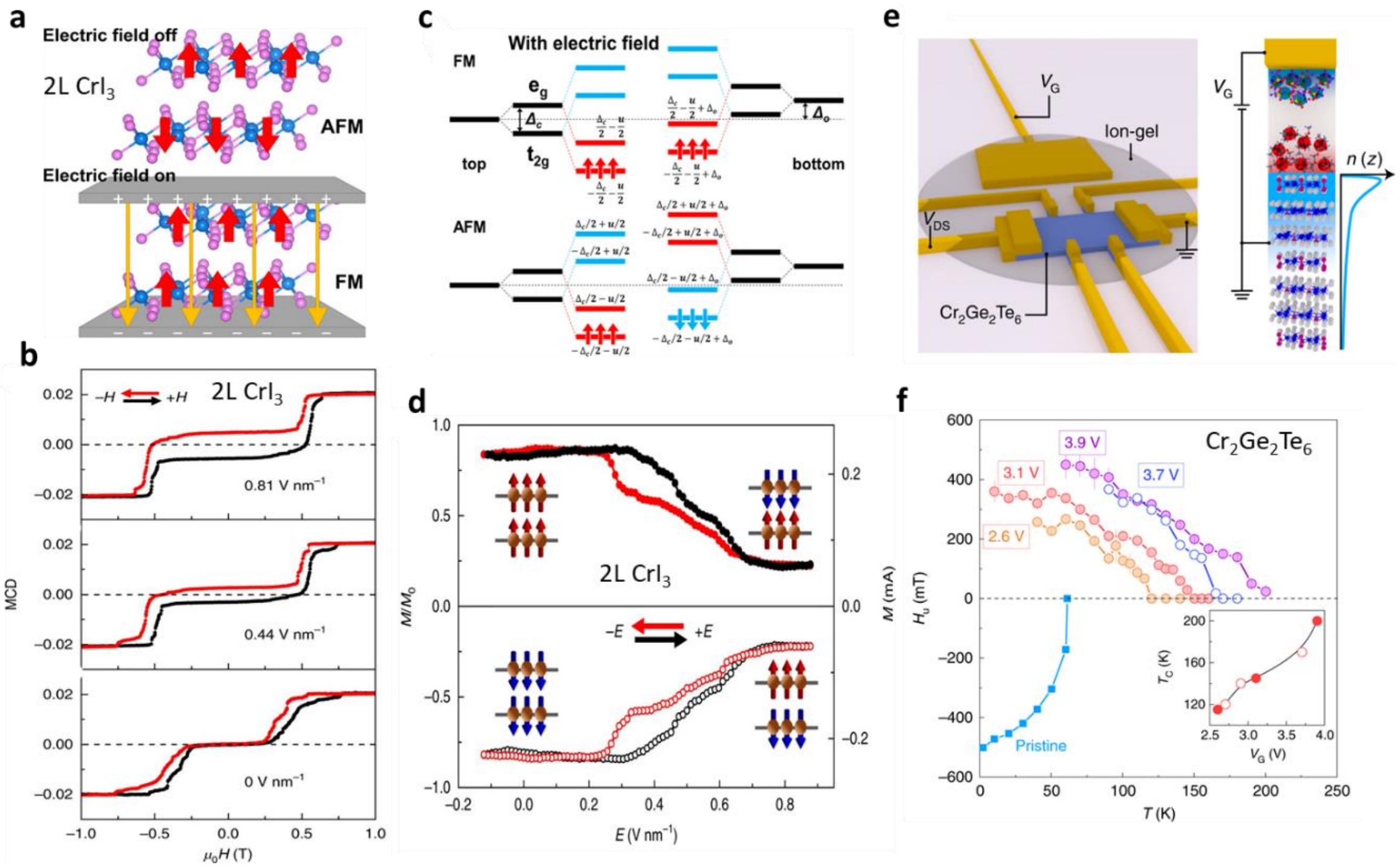

**Figure 6.** (**a**) The structure of bilayer $CrI_3$ with AFM- and FM-aligned spins when the electric field is OFF and ON, respectively; panel (a) reproduced with permission from Ref. [110]. (**b**) Magnetic field-dependent MCD curves for bilayer $CrI_3$ under different bias voltages; panel (b) reproduced with permission from Ref. [112]. (**c**) Schematic diagrams for orbital interactions based on the tight-binding theorem for $CrI_3$ with (ON) and without (OFF) electric field; panel (c) reproduced with permission from Ref. [110]. (**d**) Electric switching of magnetic order for bilayer $CrI_3$; panel (d) reproduced with permission from Ref. [113]. (**e**) The structure of an electric double-layer transistor device using $Cr_2Ge_2Te_6$ and (**f**) changes in magnetic anisotropy energy under different bias voltages with inset showing the bias voltage dependence of Curie temperature of $Cr_2Ge_2Te_6$. Panels (e,f) reproduced with permission from Ref. [117].

Of particular interest is vdW-layered $CrI_3$, whose long-range magnetic ordering has recently been experimentally realized in the monolayer through mechanical exfoliation of its bulk crystal [36]. Bulk $CrI_3$ undergoes a structural transition from a high-temperature monoclinic structure with a C2/m space group (above ~210 K) to a low-temperature rhombohedral structure with R3 symmetry (below ~210 K) [6,28]. In each vdW layer, one Cr atom is surrounded by six I-atoms, forming an edge-sharing octahedral cage. These Cr atoms are arranged within a standard honeycomb lattice (Fig. 6a). The oxidation state of Cr is 3+, leaving three 3d electrons unpaired in $CrI_3$ resulting in a magnetic moment of $3\mu_B$/Cr. It is the presence of the I-octahedral crystal field that splits the Cr-3d orbitals into triplet $t_{2g}$ and doublet $e_g$ manifolds (Fig. 6b). As a result, the residual electrons fill up the $t_{2g}$ orbitals of the spin-up channel, opening an energy gap to decrease the total energy. Bulk $CrI_3$ exhibits ferromagnetic order below ~60 K with a strong out-of-plane magnetic anisotropy. In atomically thin $CrI_3$, spins are ferromagnetically coupled within the layer and antiferromagnetically coupled between the layers, leaving the magnetic behavior of the system to vary depending upon the number of layers [6,20,24,28]. Specifically, FM or AFM order occurs in accordance with an odd or even number of layers, respectively [36]. For instance, $CrI_3$ is FM in the monolayer but AFM in the bilayer structure. While the bulk $CrI_3$ is an insulator, its 2D counterparts (monolayer, bilayer, or a few layers) are often considered “intrinsic” magnetic semiconductors. In such 2D semiconductors, the interlayer exchange interactions are rather weak and susceptible to external electrical perturbation, enabling electric field modulation of magnetic functionalities in 2D $CrI_3$ [6,28].

Upon electric gating, charge carriers will accumulate resulting in hole (electron) doping into $CrI_3$ that can alter its magnetic ordering [110-113]. For the case of monolayer $CrI_3$, the majority spins for both conduction and valence bands contribute to the electronic states near the band edges, increasing (decreasing) the coercivity, saturation magnetization, and Curie temperature upon hole (electron) doping [113]. For bilayer $CrI_3$, an exotic vdW semiconducting antiferromagnet with a Néel temperature of ~45 K, its magnetic state can be electrically modulated. DFT calculations performed by Xu *et al.* show that the application of an electric field can induce an interlayer AFM-FM phase transition in bilayer $CrI_3$ with weak interlayer versus intralayer exchange coupling (Fig. 6a) [110]. Combined DFT calculations and Monte Carlo simulations performed by Lei *et al.* also reveal the strong magnetoelectric (ME) effect in $CrI_3$ bilayers [111]. This ME response can be tuned by varying internal displacement fields and carrier densities through electric gating. The calculated dependence of magnetization on gate voltage shows a strong temperature-dependent ME response in the bilayer $CrI_3$ at charge neutrality. All this suggests the possibility for electric field modulation of magnetism in vdW AFM materials like bilayer $CrI_3$ by electrostatic doping. Indeed, using magnetic circular dichroism (MCD) microscopy, Jiang *et al.* demonstrated that the application of a sufficiently large electric field can create an interlayer potential difference, resulting in a large linear ME effect (Fig. 6c) [112]. The authors also observed complete and reversible electrical switching between the interlayer AFM and FM states in the vicinity of the interlayer spin-flip transition and have attributed this effect to the electric field dependence of the interlayer exchange bias. Using magneto-optic Kerr effect (MOKE) microscopy, Huang *et al.* also demonstrated voltage-controlled switching between AFM and FM states at fixed magnetic fields near the AFM-FM transition in bilayer $CrI_3$ (Fig. 6d) [113]. The time-reversal pair of layered AFM states at zero magnetic field exhibited spin-layer locking, causing a linear dependence of the MOKE signals on gate voltage with opposite slopes. The voltage-assisted magnetic switching and the linear ME effect have recently been shown theoretically and experimentally in twisted bilayers of $CrI_3$ [114,115]. The presence of Moiré patterns was suggested as evidence of modulated magnetic ground states and hence the ME effect. These findings revealed a wide variety of exciting magnetoelectric phenomena in vdW AFM materials that can be utilized for applications in electrically controlled high-density magnetic memory, sensors, and spintronics. Nonetheless, the exact mechanism behind electronic control of magnetism in bilayer $CrI_3$ is not fully understood, since the applied gate voltage induces both linear

ME and doping effects simultaneously. Depending on the type of AFM order, effects of electric field on the magnetic state and the ME response in AFM bilayer systems may vary and need to be studied further [116].

$Cr_2Ge_2Te_6$ is another system of interest for exploring electric control of low-dimensional magnetism and it has been extensively studied since long-range FM order was experimentally realized in bilayer form in 2017 [37]. Within its monolayer, each Cr atom is surrounded by six Te atoms with the adjacent Cr atoms connecting to form a honeycomb lattice via their ligands, with added Ge−Ge dimers. Stacking monolayers in the ABC sequence form its bulk counterpart. Like $CrI_3$, the oxidation state of Cr is 3+ in $Cr_2Ge_2Te_6$, leaving three Cr-3d electrons unpaired resulting in a local magnetic moment of $3\mu_B$/Cr [6,28]. The magnetism of bulk $Cr_2Ge_2Te_6$ is determined by the competition between superexchange FM interactions of adjacent Cr atoms mediated by Te atoms and the direct exchange AFM interaction between the nearest Cr atoms. While the AFM interaction is rather weak due to the large Cr−Cr distance (~4 Å), intralayer FM interaction is strong due to significant *p-d* hybridization. The interaction between layers (or the interlayer interaction) is also ferromagnetic. Owing to intralayer and interlayer FM interactions, bulk $Cr_2Ge_2Te_6$ undergoes a FM transition at ~68 K with an out-of-plane magnetic anisotropy [37]. Long-range FM order is preserved in the bilayer but with a reduced Curie temperature of 30 K and a rather weak magnetic anisotropy. While bulk $Cr_2Ge_2Te_6$ is an insulator, its bilayer counterpart behaves as a semiconductor whose magnetic functionalities can be modulated by electric means [117-120]. DFT calculations performed by Sun *et al.* show that the application of an external electric field to bilayer $Cr_2Ge_2Te_6$ can modulate its magnetic properties (Curie temperature, saturation magnetization, and magnetic anisotropy) due to a change in the effective band structure [117]. The electric field can also induce a semiconductor-metal transition and redistribute charges and spins between the two layers. Using an electric double-layer transistor device (Fig. 6e),

Verzhbitskiy *et al.* experimentally demonstrated the application of a gate voltage to a ~20 nm-thick $Cr_2Ge_2Te_6$ can increase its Curie temperature, saturation magnetization, and switch the magnetic anisotropy (Fig. 6f) [118]. Upon electric gating, the ferromagnetic region expanded up to 200 K (compared to $T_C$ ~ 68 K for bulk (undoped) $Cr_2Ge_2Te_6$) and a noticeable change in the magnetic easy axis from out-of-plane to in-plane appears to occur. A similar effect was also reported by Zhou *et al.* [119] and Wang *et al.* [120] for few-layered $Cr_2Ge_2Te_6$ samples. It can be seen in the inset of Fig. 6f that the Curie temperature can be easily tuned by varying gate voltages. It has been suggested that the large enhancement in $T_C$ resulted from the double-exchange FM interaction activated by introducing a significant number of electrons into the material upon electric gating. Due to this doping of electrons, some of the $Cr^{3+}$ ions are converted into $Cr^{2+}$ ions. As electrons hop between $Cr^{2+}$ and $Cr^{3+}$ ions via Te atoms, this process occurs simultaneously in accordance with the DE mechanism. This DE FM interaction is dominant over the SE interaction that governs the insulating FM state of $Cr_2Ge_2Te_6$. This probably causes the magnetic easy axis to switch from the out-of-plane to the in-plane direction when $Cr_2Ge_2Te_6$ is electrically gated.

*2.2.2.2. High temperatures*

While the 2D intrinsic semiconductors $CrI_3$ and $Cr_2Ge_2Te_6$ show electrically tunable magnetic properties including a large ME effect, these materials are restricted to operate at low temperatures (below 200 K) [113-115,118-120]. Practical applications of spin-based devices usually require the electrical control of magnetism at or above room temperature. 2D ferromagnetic semiconductors with high Curie temperatures are therefore highly sought.

Recent studies have shown that the introduction of magnetic transition metal atoms into semiconducting 2D-TMDs $TX_2$ ($T$ = V, Fe, Cr; $X$ = S, Se, Te), such as V-doped $WS_2$ [50], V-doped $WSe_2$ [51,52], and Fe-doped $MoS_2$ [53] monolayers, permits long-range ferromagnetic order that can be induced at room temperature (see, for example, Fig. 7a,b for the 4 at.% V-doped $WSe_2$ monolayer [51]). These materials form a novel class of 2D DMSs [50-53]. The origin of the ferromagnetic order in these 2D-TMD systems was explained by the RKKY mechanism, where free holes promote FM interaction between V atoms. Zhang *et al.* reported that p-type V-doped $WS_2$ monolayers possess tunable ferromagnetism at room temperature [50]. By replacing W, having six valence electrons, with V, having five valence electrons, an electron deficiency is created in V-$WS_2$ that eventually becomes a p-type dominant semiconductor. Unlike diamagnetic pristine $WS_2$, ferromagnetism emerges and becomes enhanced with V doping in monolayers of V-$WS_2$, which is optimal at ~2 at.% V but quenched for higher V concentrations due to orbital hybridization at too-close vanadium-vanadium distances. In a similar manner, strong and tunable ferromagnetism has been achieved in V-doped $WSe_2$ monolayers with a 4 at.% V-doping concentration showing the largest saturation magnetization (Fig. 7b) [51]. The uniform distribution of V atoms in these TMDs is crucial to establishing long-range FM order at high temperatures, with the optimal magnetization achievable at high V concentrations. However, heavy doping with V atoms can cause these magnetic dopants to agglomerate, inducing AFM couplings or spin disordering at the expense of the FM couplings. As a result, the net magnetization of the material is drastically reduced or even vanishes [50-52]. In such heavily doped 2D-TMD systems, photoluminescence (PL) features are strongly suppressed, and their semiconducting characteristics are significantly altered [50,52]. Current efforts are devoted to enhancing and modulating the magnetic and magneto-optic responses of 2D DMSs through control of external stimuli such as electric fields [52,55] and light [54].

To demonstrate the electric field modulation of ferromagnetism in V-doped $WSe_2$ monolayers, Yun *et al.* performed magnetic force microscopy (MFM) measurements on a 0.1% V-doped $WSe_2$ monolayer under different gate voltages ranging from -10 to 20 V (Fig. 7c) [52].

Significant changes in phase contrast of the magnetic domains of this sample appeared to occur when a positive gate voltage was applied (Fig. 7d), unlike the application of negative gate voltages (-10 V and -5 V). Interestingly, the phase deviation increased with increasing gate voltage from 0 to 15 V, indicating enhanced FM interactions or magnetization due to hole doping. The phase deviation dropped when the gate voltage exceeded 15 V; however, the reason for this drop was not well understood. To complement these experimental findings, Duong *et al.* performed ab initio calculations of injected charge (hole, electron) concentration-dependent exchange energy and magnetic moment for a V-doped $WSe_2$ monolayer [55]. The calculated results show that while electron injection shifts the Fermi level towards the conduction band edge, hole injection shifts the Fermi level deeper inside the valence band. Both the exchange energy and the total magnetic moment become larger when holes are injected, while an opposite trend is found for electron injection (Fig. 7e). It is worth noting here that in the V-doped $WSe_2$ monolayer, the V atom couples antiferromagnetically to the nearest W sites and ferromagnetically to the distant W sites. The introduction of charge carriers mediates this interaction, where increasing hole carriers results in an enhanced magnetic moment at the V site. Additionally, the magnetic moment near the W sites flips from antiferromagnetic to weakly ferromagnetic, which combines with the increased magnetic moment at the V site, resulting in enhanced long-range ferromagnetism. On the other hand, electron doping severely reduces the total magnetic moment of the V and distant W sites, shunting long-range ferromagnetic order (Fig. 7e). The slight reduction of the total magnetic moment at the highest hole density ($2h^+$) has been attributed to the screening effect of hole carriers.

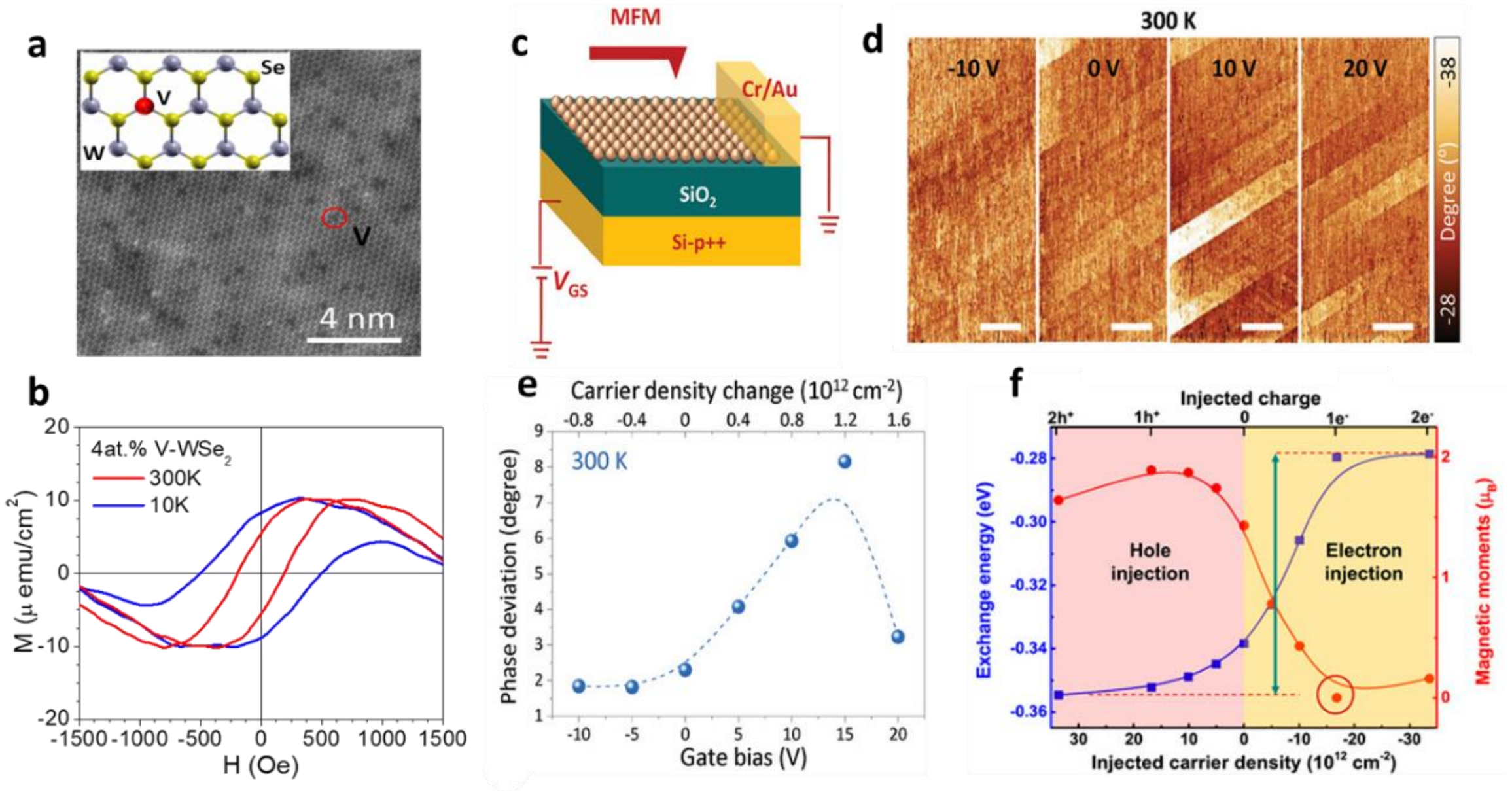

**Figure 7.** (**a**) TEM image of a 4 at.% V-doped $WSe_2$ monolayer with its crystal structure shown in the inset. (**b**) Magnetic hysteresis (M-H) loops of the 4 at.% V-doped $WSe_2$ monolayer taken at 10 and 300 K; panels (a,b) reproduced with permission from Ref. [51]. (**c**) Schematic of gate voltage-dependent MFM measurements. (**d**) Gate voltage-dependent MFM images taken at 300 K for the 0.1 at.% V-doped $WSe_2$ monolayer; panels (c,d) reproduced with permission from Ref. [52]. (**e**) Exchange energy and net magnetic moment of the V-doped $WSe_2$ monolayer with different carrier (hole, electron) doping densities. Panel (e) reproduced with permission from Ref. [55].

The effect of electric gating (hole injection or hole doping) on FM ordering in V-doped $WSe_2$ monolayers is rather similar to that of light irradiation observed by Jimenez *et al.* for the same system [54], as well as for the V-doped $WS_2$ monolayer [50]. In the latter, upon light irradiation, the absorbed photons generate electron-hole pairs. The electrons are captured by electron-deficient V sites, which generate an imbalance in the carrier population (extra holes are introduced into the V-$WSe_2$ monolayer), resulting in an enhanced magnetic moment [54,121]. The total magnetic moment has been observed to increase with increasing light intensity (hole density) and to saturate at larger light intensities [54]. A similar hole density-dependent magnetic moment trend has been observed for both electric gating and light irradiation effects [52,54]. These findings provide solid evidence for hole-mediated long-range ferromagnetism in magnetically-doped TMD monolayers.

While both DFT calculations and MFM measurements reveal negligible change in the magnetism of pristine $WSe_2$ monolayer upon electric gating, Nguyen *et al.* recently reported gate-tunable magnetism in multilayers of $WSe_2$ via resonant Se vacancies [121]. Photocurrent measurements under different gate biases were performed to probe Se-vacancy spin states in the sample through light-induced conversion of unoccupied states to partially occupied states. It should be noticed that defective magnetism or defect engineering of magnetism in TMD systems is not uncommon. A large body of work has shown that ferromagnetism can be induced in TMDs such as $MoS_2$ and $WSe_2$ by engineering defects (e.g., transition metal (Mo, W) or chalcogen (S, Se) vacancies) [122]. However, exact configurations and concentrations of defects created in most of the investigated TMDs (during growth or heat treatment) are unknown. Therefore, the physical origin of defect-induced ferromagnetism in these systems has remained elusive. The case becomes even more complex when magnetic transition metal atoms are incorporated into TMD lattices to form DMSs, such as V-doped $WSe_2$ monolayers [50], because decoupling interrelated contributions to the total magnetic moment from the defect- and dopant-induced magnetic moments in these magnetically-doped TMDs is a very challenging task [50-53,122,123]. While the observation of gate-tunable magnetism via chalcogen vacancies in TMDs is very promising for applications in spintronics, further studies are needed to fully understand how chalcogen/transition metal vacancies are formed, effects of these defect configurations, and vacancy-dopant magnetic couplings on the gate-modulated magnetic phenomena in both pristine and doped 2D-TMD systems. In this regard, Co-doped ZnO vdW semiconducting monolayers whose ferromagnetism has recently been reported at room temperature are also of potential interest for the exploration of gate-tunable magnetic and magneto-optic devices [124]. This is feasible since the electric field modulation of ferromagnetism has already been demonstrated in a back-gated Mn-doped ZnO nanowire field-effect transistor, where confinement and dimensionality effects were found significant [125].

### *2.2.3. Heterostructures*

Of practical interest, vdW materials allow wide flexibility and integration with each other. Stacking different vdW materials has been reported to create novel heterostructures with atomically sharp interfaces and properties that would otherwise be absent in their individual components [6,20,21,27,28]. In addition to their outstanding magnetic properties, 2D vdW intrinsic magnets ($Fe_3GeTe_2$, $CrI_3$, and $Cr_2Ge_3Te_6$) can also be interfaced with other vdW or non-vdW materials to create heterostructures with new and novel functionalities [20,24,28]. We discuss below how the magnetic, magneto-optic, and valleytronic properties of such heterostructures arise and can be manipulated by electric means and whose applications might span over a wide spectrum ranging from spintronics to spincaloritronics and valleytronics.

#### *2.2.3.1. $Fe_3GeTe_2$-based heterostructures*

Owing to its high Curie temperature and strong out-of-plane magnetic anisotropy, $Fe_3GeTe_2$ has been integrated with either vdW or non-vdW materials to create heterostructures with enhanced magnetic functionalities including the emerging exchange bias (EB) effect (or the exchange anisotropy) attributable to their interfacial magnetic exchange couplings [126-133]. A comprehensive review of EB and interface-related effects in the $Fe_3GeTe_2$ (or $Fe_5GeTe_2$)-based vdW heterostructures has recently been made by Phan *et al.* [35]. In this context, the electric field manipulation of the EB effect or exchange anisotropy is desirable for spintronic devices such as those based on spin-orbit-torque (SOT) technology [133].

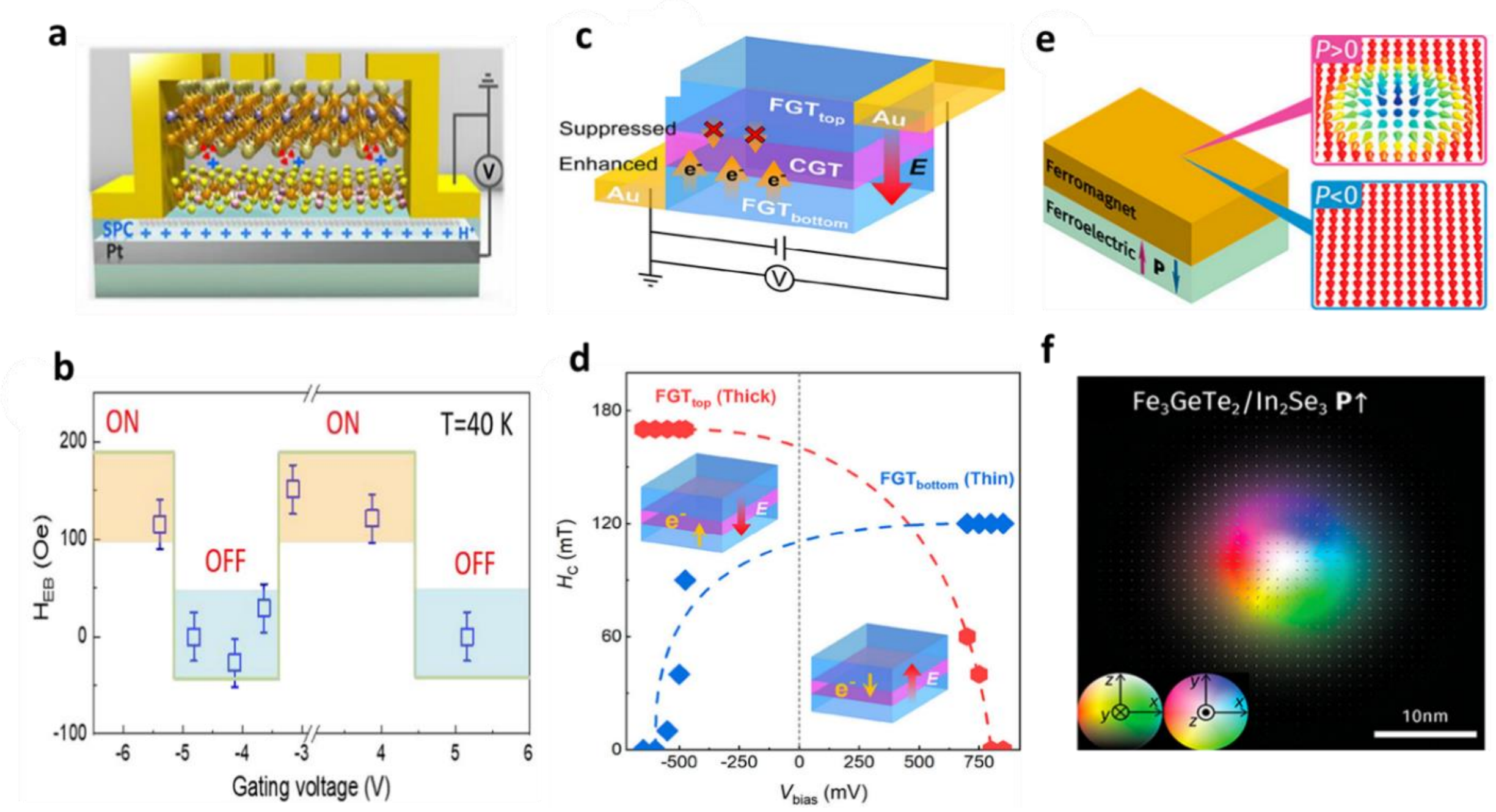

**Figure 8. (a)** The structure of a gating device based on an $Fe_5GeTe_2/FePS_3$ heterostructure. **(b)** Exchange bias field ($H_{EB}$) of $Fe_5GeTe_2/FePS_3$ changing with gate voltage; panels (a,b) reproduced with permission from Ref. [132]. (**c**) Schematic of an FGT/CGT/FGT MTJ and the effect of directional electric fields on charge transfer. (**d**) Extracted bias-voltage-dependent coercive field ($H_C$) of the top and bottom FGT electrodes; panels (c,d) reproduced with permission from Ref. [133]. **(e)** Schematic of a $Fe_5GeTe_2/In_2Se_3$ heterostructure, where the switching of electric polarization ($In_2Se_3$) can reversibly create and annihilate magnetic skyrmions in $Fe_5GeTe_2$. **(f)** Atomistic spin-dynamics modeling result of the $Fe_5GeTe_2/In_2Se_3$ heterostructure for positive ferroelectric polarization. Panels (e,f) reproduced with permission from Ref. [136].

Albarakati *et al.* recently demonstrated the capacity of tuning the EB effect in a $Fe_5GeTe_2/FePS_3$ FM/AFM vdW heterostructure by applying a solid protonic gate [132]. They showed that the EB field could reach up to 23% of the coercivity and the blocking temperature varied from 30 to 60 K under various gate voltages. Figures 8a and b illustrate the structure of an $Fe_5GeTe_2/FePS_3$-based device and how the EB effect or the EB field ($H_{EB}$) of the $Fe_5GeTe_2/FePS_3$ heterostructure was electrically controlled at a given temperature (e.g., 40 K) below the blocking temperature (~60 K).

The EB effect in the $Fe_5GeTe_2/FePS_3$ heterostructure has been attributed to proton intercalation [132]. Upon electric gating, proton intercalation increases, which decreases the interfacial magnetic coupling between $Fe_5GeTe_2$ and $FePS_3$ layers hence altering the EB effect. It has been suggested that proton intercalation can not only mediate the magnetic exchange coupling but also alter AFM configurations within the $FePS_3$ layer. This approach is quite promising and can be applied to $Fe_3GeTe_2$-based heterostructures whose exchange anisotropy is found significant [126-131]. Of importance is the air sensitivity of $Fe_3GeTe_2$ since air can cause both surfaces of an $Fe_3GeTe_2$ film to oxidize, resulting in two oxidized layers with different magnetic configurations (often considered as AFM) with which the remaining (non-oxidized) FM layer can couple to induce EB effects [129,130]. A careful characterization of spintronic devices that utilize the EB effect and electric gating is therefore needed.

Recall that in such vdW heterostructures, charge transfer may play a crucial role in mediating magnetic and transport properties across the interfaces [35,133]. An interesting question arises: *Can this charge transfer-mediated interfacial magnetism can be modulated by electric means?* Recently, Wang *et al.* performed magnetoresistance measurements on a $Fe_3GeTe_2/Cr_2Ge_2Te_6/Fe_3GeTe_2$ heterojunction (Fig. 8c) and demonstrated the possibility of controlling the magnetic functionality (e.g., coercivity) of the top or bottom $Fe_3GeTe_2$ electrode (Fig. 8d) through gate voltage manipulation of directional charge transfer that occurs across the $Fe_3GeTe_2/Cr_2Ge_2Te_6$ interface [133]. First-principles calculations show that the directional charge transfer from the metallic $Fe_3GeTe_2$ layer to the semiconducting $Cr_2Ge_2Te_6$ layer modifies the magnetic anisotropy energy of $Fe_3GeTe_2$, causing a dramatic suppression of coercive field. This study suggests a new perspective on manipulating the magnetic properties of ferromagnetic electrodes in vdW magnetic heterojunctions through a voltage-controllable charge transfer mechanism.

To effectively explore an EB effect-based spintronic device, the area comprising the FM/AFM interface should be sufficiently large. However, optimization of device size and structure require this area to be as small as possible [35]. This technical challenge can be overcome by considering nonlocal manipulation of magnetism in FM/AFM vdW heterostructures. Through MOKE measurements, Dai *et al.* demonstrated the magnetic functionality of $Fe_3GeTe_2$ can be nonlocally manipulated by partially forming its interface with an AFM $MnPS_3$ and passing an ac current through its heterostructure [134]. It was shown that the coupling of a small piece of $MnPS_3$ (~40 $\mu m^2$) with FGT not only enhanced the coercive field but also induced the EB effect over the entire flake of $Fe_3GeTe_2$ (~2000 $\mu m^2$). It should be recalled that in an MRAM unit, an AFM layer is used to provide an EB field to realize controllable magnetization reversal. In this context, it is possible to design a magnetic device that employs a small FM/AFM area-based tip to control the magnetism of an entire $Fe_3GeTe_2$ layer while reading information from the remaining area of $Fe_3GeTe_2$. Such a vdW magnetic device appears to have a much simpler design and fabrication compared to conventional non-vdW magnetic devices [10,23].

As mentioned above, magnetic skyrmions, which are often induced by DMI, are an interesting platform for the development of modern spintronics [59,60]. Electric control of magnetic skyrmions is particularly ideal for application in low-power memory devices [60].

Recent works have revealed that magnetic skyrmions can be created in $Fe_3GeTe_2$ and its heterostructures [70-79]. Using photoemission electron microscopy and Lorentz transmission electron microscopy, Yang *et al.* showed that magnetic field-free Néel-type magnetic skyrmions could be formed and stabilized in $Fe_3GeTe_2$ when it was interfaced with $(Co/Pd)_n$ [72]. Wu *et al.* showed, via Hall effect measurements and magnetic force microscopy, that two groups of magnetic skyrmions could be formed at a vdW interface comprising two 2D ferromagnets, $Fe_3GeTe_2$ and $Cr_2Ge_2Te_6$ [79]. The authors also found that the magnetic skyrmions could even persist in zero magnetic field applied to $Fe_3GeTe_2/Cr_2Ge_2Te_6$ heterostructures. These experimental observations are supported by micromagnetic simulations, showing a new perspective for controlling multiple skyrmion states for use in vdW magnetic heterostructure-based skyrmionic devices. Given the fact that the interfacial magnetic anisotropy in these heterostructures can be modulated by electric fields, it would be very interesting to investigate how these magnetic skyrmions can be stabilized and manipulated using this voltage-controlled interfacial magnetic anisotropy [135]. Based on DFT and atomistic spin-dynamics modeling, Huang *et al.* recently showed that DMI and magnetic skyrmions in a monolayer of $Fe_3GeTe_2$ can be manipulated by the ferroelectric polarization of an adjacent 2D $In_2Se_3$, which is a ferroelectric vdW material [136]. In case of the $Fe_3GeTe_2/In_2Se_3$ heterostructure, it is possible to create and annihilate magnetic skyrmions by reversing the ferroelectric polarization of $In_2Se_3$ that eventually alters the DMI magnitude (Fig. 8e,f). Calculations also show that ferroelectric switching can change the sign of DMI and cause a skyrmion chirality reversal. These predictions are very exciting but need to be experimentally validated.

In addition to the roles played by charge and spin, valleys, which refer to the energy extrema in the electronic band structure of a semiconductor like a TMD monolayer (e.g., $WSe_2$, $WS_2$, and $MoS_2$), are considered a new degree of freedom of electrons [137-140]. Valleytronics, which combines valley and electronics, has become a hot topic of research in recent years [137-146]. Electrical manipulation of the valley degree of freedom in the 2D-TMDs is challenging but essential for developing valleytronic devices [144-146]. Interfacing a 2D-TMD with a magnetic material has been shown to enhance valley Zeeman splitting in the 2D-TMD through the magnetic proximity effect (MPE) induced by the ferromagnetic layer [141,145-147]. This FM layer can also inject spin-polarized carriers into the TMD layer to realize valley-dependent polarization. Instead of using non-vdW materials Ga(Mn)As and NiFe as ferromagnetic contacts in 2D TMD-based heterojunctions, the use of the 2D vdW ferromagnet $Fe_3GeTe_2$ minimizes the lattice mismatch issue and yields an atomically flat vdW interface with monolayer $WSe_2$, resulting in the gate-tunable valleytronics in this heterostructure [147]. Kim *et al.* observed a large enhancement of interfacial perpendicular magnetic anisotropy at low temperatures in thinned crystals of $Fe_3GeTe_2$ when interfaced with monolayer $WSe_2$ [148]. The origin of this enhancement was not well understood though. It will thus be interesting to investigate how MPE impacts the interfacial magnetic anisotropy in the $Fe_3GeTe_2/WSe_2$ heterostructure, and whether charge transfer across the $Fe_3GeTe_2/WSe_2$ interface mediates this interfacial magnetic anisotropy. Since the magnetic functionality of FGT can be manipulated by both magnetic and electric fields, it is possible to employ these external fields to control the interfacial magnetic anisotropy and hence the valley state in the $Fe_3GeTe_2/WSe_2$ system, which appears to be promising for applications in spin valleytronics.

#### *2.2.3.2. $CrI_3$-based heterostructures*

As described above, it is possible to electrically control the magnetic order in bilayer $CrI_3$ via the linear ME effect [112-115]. However, the electric control of magnetism can be achieved only near the magnetic field-driven AFM to FM transition in bilayer $CrI_3$, and this approach is not applicable to exploring monolayer $CrI_3$, which belongs to the class of centrosymmetric materials. To overcome this, Jiang *et al.* formed $CrI_3$/graphene vertical heterostructures and demonstrated the capacity to manipulate the magnetic properties of both monolayer and bilayer $CrI_3$ via electrostatic doping [149]. It was shown that hole/electron doping could significantly alter the saturation magnetization, coercive force, and Curie temperature of the $CrI_3$ monolayer. Interestingly, electron doping above ~$2.5 \times 10^{13}$ $cm^{-2}$ promoted an AFM to FM transition in bilayer $CrI_3$ without the need of an external magnetic field. These findings demonstrate the possibility of using small gate voltages to switch the magnetization in bilayer $CrI_3$. Using dual-gated graphene/$CrI_3$/graphene tunnel junctions to fabricate spin tunnel field-effect transistors, Jiang *et al.* showed that the tunnel conductance depended on the magnetic order in the $CrI_3$ tunnel barrier and that the applied gate voltage could effectively switch the tunnel barrier between AFM and FM states near the spin-flip transition [150]. These spin transistors achieved a high-low conductance ratio of up to 400%, which is promising for use in non-volatile memory devices.

While most studies focused on the static magnetic properties of vdW magnets and heterostructures, electric field control of their dynamic magnetic properties has been less explored. However, Zhang *et al.* recently showed that the spin-wave dynamics of an antiferromagnetic $CrI_3$ bilayer layer in the $CrI_3$/$WSe_2$ heterostructure can be electrically modulated using an ultrafast optical pump/MOKE probe technique [151]. The presence of monolayer $WSe_2$ with strong excitonic resonance has been shown to enhance the optical excitation of spin waves. The authors have demonstrated the possibility of tuning antiferromagnetic resonances by tens of gigahertz through electrostatic gating, which creates a new outlook for applications in ultrafast data storage and processing.

On the other hand, it has been shown that the presence of 2D $CrI_3$ at the $CrI_3/WSe_2$ interface can induce the MPE on the $WSe_2$ monolayer which, in turn, alters its valley splitting state [146]. Indeed, Li *et al.* showed that the valley degeneracy in the $WSe_2$ monolayer was lifted by the MPE of the $CrI_3$ layer, and that the valley polarization in this TMD monolayer could be switched by electrical gating [146]. The application of a sufficiently high bias voltage could switch the magnetization of the top $CrI_3$ layer (relative to the bottom layer) adjacent to the $WSe_2$ layer, leading to the switching of the valley polarization in the $WSe_2$ layer. These experimental findings are also complemented by first-principles calculations performed by Zollner *et al.* [152]. In the latter, the authors predicted proximity exchange in monolayer $MoSe_2$ or $WSe_2$ due to the presence of the FM monolayer $CrI_3$, which can be tuned by electric gating. DFT calculations performed by Zhang *et al.* also show the important role of W-Cr superposition for the large valley splitting in the monolayer $CrI_3/WSe_2$ heterostructure, which can also be enhanced or tuned in the trilayer $CrI_3/WSe_2/CrI_3$ heterostructure with appropriately aligned layers [153]. This is in line with the theoretical work of Ge *et al.* that shows that the valley splitting of monolayer $Wse_2$ can be enhanced in twisted $Wse_2/CrI_3$ heterostructures [154]. These predictions, however, remain to be observed by experiments.

Like $Fe_3GeTe_2$, using the ferroelectric polarization of a ferroelectric vdW material interfaced with 2D $CrI_3$ has been proposed to control the magnetic state of $CrI_3$ [155,156]. First-principles calculations have suggested that interlayer magnetism of bilayer $CrI_3$ can be switched between the AFM and FM state by ferroelectric polarization reversal of $\alpha$-$In_2Se_3$, which is a ferroelectric vdW material [155]. In this case, the polarization of $\alpha$-$In_2Se_3$ has been shown to mediate the interlayer magnetic coupling between $CrI_3$ monolayers, whose mechanism is driven by the direct interaction of adjacent I atoms. In a similar fashion, the ferroelectric polarization of ferroelectric $Sc_2CO_2$ at the $CrI_3/Sc_2CO_2$ interface has also been theoretically shown to mediate the interlayer magnetic coupling between $CrI_3$ layers, leading to switching between AFM and FM states [156]. The calculations reveal that the strong ME effect arises from a dramatic change of the band alignment induced by the strong built-in electric polarization in $Sc_2CO_2$ and the subsequent change of the interlayer magnetic coupling of bilayer $CrI_3$.

When a FM monolayer $CrI_3$ is brought to interface with a ferroelectric $Sc_2CO_2$ layer, Zhao *et al.* theoretically showed that the magnetism of the monolayer can be controlled by the ferroelectric polarization reversal of $Sc_2CO_2$ (polarized states, P↑ and P↓) [157]. It has been shown that while the positive ferroelectric polarization (P↑) of $Sc_2CO_2$ does not change the semiconducting nature of monolayer $CrI_3$, the positive ferroelectric polarization (P↓) of $Sc_2CO_2$ makes the monolayer $CrI_3$ half-metallic. This demonstrates a new perspective for nonvolatile electric control of 2D vdW ferromagnets in half-metal-based nanospintronics.

From another standpoint, Sun *et al.* reported the observation of a complete Néel-type skyrmion-bimeron-ferromagnet phase transition in the $WTe_2$/$CrCl_3$ heterostructure [158]. This transition is mediated by a perpendicular magnetic field and is driven by the competition between the out-of-plane magnetocrystalline anisotropy and magnetic dipole-dipole interaction. The authors proposed the possibility of controlling these skyrmions using the ferroelectric polarization of a ferroelectric $CuInP_2S_6$ monolayer in contact with the $CrCl_3$ layer. Interestingly, Hu *et al.* recently predicted that $CuInP_2S_6$ itself can change its valley state from paravalley to ferrovalley when interfaced with antiferromagnetic $MnPS_3$ due to the MPE caused by the $MnPS_3$ layer [159].

### *2.2.3.3. $Cr_2Ge_3Te_6$-based heterostructures*

To enhance the electric field-modulated magnetic functionality of 2D $Cr_2Ge_2Te_6$, its heterostructures with other materials have been developed [160-164]. Ostwall *et al.* showed, via anomalous Hall effect measurements, that a combination of an in-plane magnetic field and a charge current flowing through a heavy metal (e.g., Ta) could be made to control the magnetization of a vdW ferromagnet $Cr_2Ge_2Te_6$ in a $Cr_2Ge_2Te_6$/Ta heterostructures [160]. In the presence of an applied 20 mT field, a low charge current density (~$5 \times 10^5$ A $cm^{-2}$) is sufficient to switch the out-of-plane magnetization of $Cr_2Ge_2Te_6$. This charge current density is much lower than that required for spin-orbit torque switching using non-vdW magnets such as CoFeB (~$2 \times 10^7$ A $cm^{-2}$).

To electrically switch the magnetization of 2D $Cr_2Ge_2Te_6$, this material has also been proposed to interface with a ferroelectric vdW material such as $In_2Se_3$ [161] or $SC_2Co_2$ [162], whose ferroelectric polarizations are utilized to control the magnetic state of $Cr_2Ge_2Te_6$. Based on DFT calculations, Gong *et al.* showed that the magnetization of the $Cr_2Ge_2Te_6$ monolayer can be switched as $In_2Se_3$ reverses its ferroelectric polarization direction [161]. On the other hand, $In_2Se_3$ becomes a switchable magnetic semiconductor due to the MPE of the FM $Cr_2Ge_2Te_6$ layer. In a similar perspective, the magnetization of the $Cr_2Ge_2Te_6$ monolayer in the $Cr_2Ge_2Te_6/Sc_2CO_2$ heterostructure has been predicted to be switchable by the ferroelectric polarization reversal of $Sc_2CO_2$ [162]. Like the case of the $CrI_3/Sc_2CO_2$ heterostructure, DFT calculations suggested that the positive ferroelectric polarization (P↑) of $Sc_2CO_2$ does not change the semiconducting nature of the $Cr_2Ge_2Te_6$ monolayer, but the negative ferroelectric polarization (P↓) of $Sc_2CO_2$ converts $Cr_2Ge_2Te_6$ into a half-metal. Interestingly, the magnetic easy axis of $Cr_2Ge_2Te_6$ has been shown to switch from in-plane to out-of-plane as the ferroelectric polarization of $Sc_2CO_2$ is reversed from the negative to positive direction. All these interesting predictions require experimental verifications.

On the other hand, the ferromagnetic polarization of 2D $Cr_2Ge_2Te_6$ has been explored to control the topological surface state of a 3-dimensional topological insulator such as $BiSbTeSe_2$. An AHE with abrupt hysteretic switching was reported for this $Cr_2Ge_2Te_6/BiSbTeSe_2$ heterostructure [163]. This effect has been attributed to Berry curvature associated with an exchange gap induced by interaction between the topological surface state and out-of-plane magnetic ordering. Interestingly, the authors have shown a new possibility of tuning AHE amplitude using gate voltages.

The combined magnetic and optic properties of 2D $Cr_2Ge_2Te_6$ have also been explored to enhance (via the MPE) and manipulate (electric gating) the valleytronic properties of semiconducting TMD monolayers. By forming a $Cr_2Ge_2Te_6$/$MoSe_2$ heterostructure (Fig. 9a,b), Zhang *et al.* demonstrated enhanced valley splitting in the $MoSe_2$ monolayer due to the MPE of the ferromagnetic $Cr_2Ge_2Te_6$ layer and control of the valley polarization of the $MoSe_2$ monolayer by electric gating (Fig. 9c-e) [164]. At a given gate voltage, the magnetic field and temperature dependences of valley polarization were found significant and were attributed to the ferromagnetic polarization of $Cr_2Ge_2Te_6$. It can be seen in Fig. 9c-d that the degree of circular polarization (DOCP) in the $Cr_2Ge_2Te_6$/$MoSe_2$ heterostructure can be controlled by both external magnetic fields and gate voltages. Figure 9f describes the mechanism of spin-polarized charge transfer that occurs in the $Cr_2Ge_2Te_6$/$MoSe_2$ heterostructure under an applied magnetic field, leading to the formation of trions. Recent DFT calculations have complemented these experimental findings, showing that the magnitude of the valley polarization increases from 3 meV for zero electric field to 9 meV for an applied electric field of 0.4 V $Å^{-1}$ in a $Cr_2Ge_2Te_6$/$MoSe_2$ heterostructure [165].

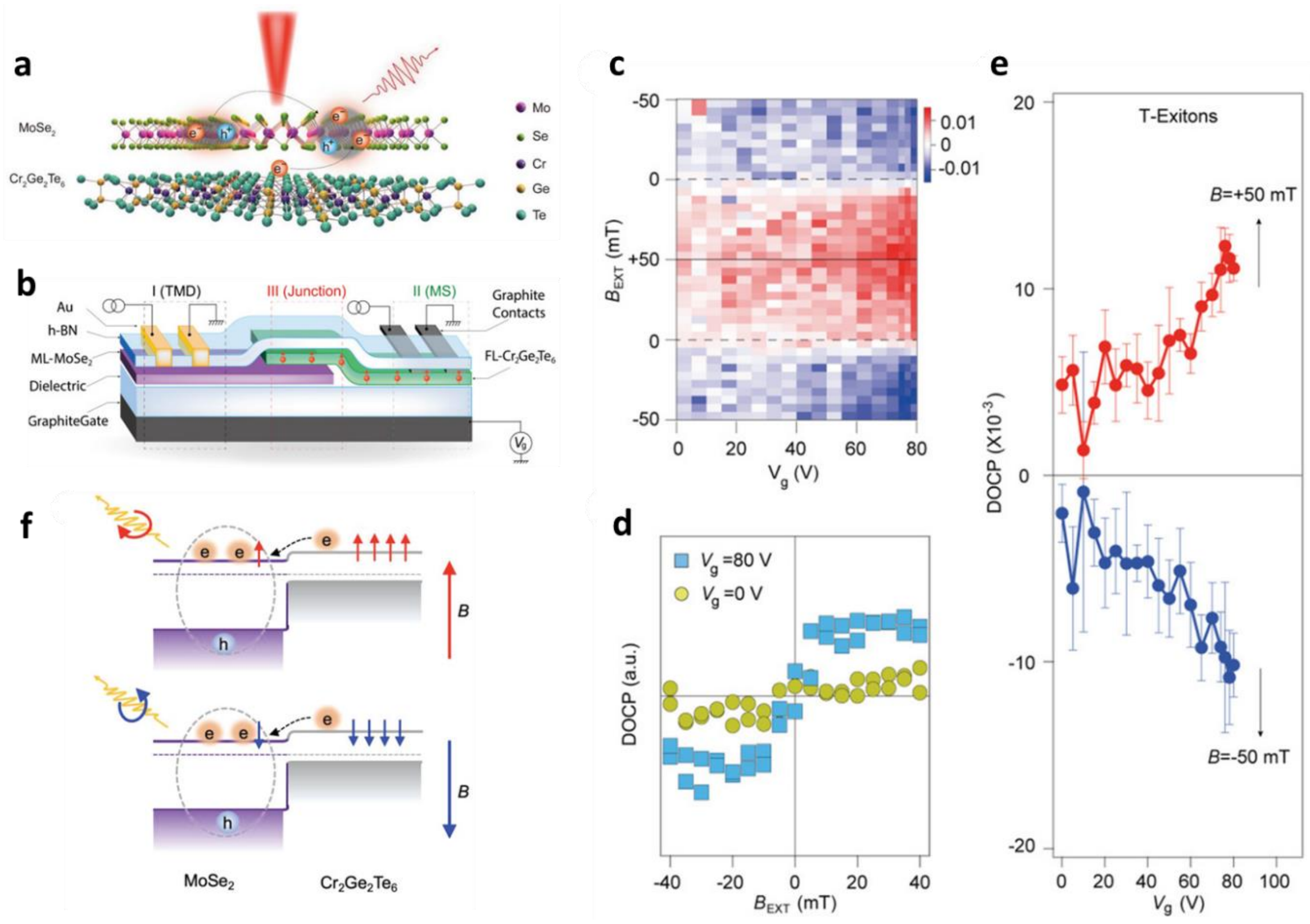

**Figure 9.** (**a**) The crystal structure of a $Cr_2Ge_2Te_6/MoSe_2$ heterostructure. (**b**) The structure of a $Cr_2Ge_2Te_6/MoSe_{2\text{-based}}$ device. (**c-d**) Electrically and magnetically tunable valley polarization in the $Cr_2Ge_2Te_6/MoSe_2$ heterostructure, with (**c**) as the color map of DOCP for trions in the heterojunction as a function of gate voltage and applied magnetic field and **(d)** is DOCP as a function of magnetic field at $V_g = 0$ V and $V_g = 80$ V. (**e**) Schematic illustration of spin-polarized charge transfer phenomena and trion formation in the $Cr_2Ge_2Te_6/MoSe_2$ heterojunction upon application of an external magnetic field. Reproduced with permission from Ref. [164].

*2.2.3.4. Transition metal dichalcogenide (TMD)-based heterostructures*

While nonmagnetic (diamagnetic) 2D-TMDs such as $WS_2$ and $MoS_2$ are semiconducting, magnetic 2D-TMDs such as $VSe_2$, $MnSe_2$, and $CrSe_2$ are metallic [40,42,45]. For spintronic applications, it is ideal to have a single material with combined ferromagnetic and semiconducting properties [9,13]. The incorporation of small amounts of magnetic dopant (V, Fe, Cr) into a semiconducting TMD monolayer has been shown to induce FM order at room temperature [50-53], however, this may lead to a strong suppression of the photoluminescent response of the material [50,57]. To overcome this, interfacing a nonmagnetic 2D-TMD semiconductor (e.g., $VS_2$, $MoS_2$, $WSe_2$) with a magnetic TMD metal ($VSe_2$, $MnSe_2$, $CrSe_2$) appears to be a promising alternative approach to creating novel 2D-TMD heterostructures as the magnetic TMD layer can magnetize the nonmagnetic 2D-TMD layer (e.g., $WSe_2$, $MoS_2$) via the MPE while preserving the semiconducting characteristic of the nonmagnetic TMD layer [40,166-169].

Recently, Marfoua and Hong performed a theoretical investigation of the effects of electric field on the Curie temperature, anomalous Hall conductivity (AHC), and anomalous Nernst conductivity (ANC) of a 1T-$VSe_2$/$MoTe_2$ heterostructure [168]. They showed that the Curie temperature of the $VSe_2$/$MoTe_2$ heterostructure can significantly increase from 270 K (the Curie temperature of the $VSe_2$ monolayer [103]) to 355 K when an electric field was applied. While a negative bias voltage could cause switching of the AHC, no AHC switching was observed under applied positive bias voltages. A large ANC of 2.3 A $K^{-1}$ $m^{-1}$ was obtained when the electric field was applied from the $VSe_2$ to the $MoTe_2$ layer, whereas it is switched to −0.6 A $K^{-1}$ $m^{-1}$ when the electric field was reversed. Theoretical calculations by Zhou *et al.* for a 1T-$VSe_2$/$MoS_2$ heterostructure-based magnetic tunnel junction (MTJ) demonstrated a substantial tunneling magnetoresistance (TMR) of 846% at room temperature [170]. Owing to the strong SOC of $MoS_2$ and its coupling with $VSe_2$, it was possible to switch the magnetization of the $VSe_2$ monolayer via the SOT mechanism. This suggests the possibility of controlling both magnetism and spin transport in the $VSe_2$/$MoS_2$ heterostructure by electric means. Since the $VSe_2$ layer has a larger work function of ~4.5 eV as compared to the $MoS_2$ layer (~4.1 eV), charge transfer likely occurs across the $VSe_2$/$MoS_2$ interface [40]. Electrons accumulate in the $VSe_2$ layer and a depleted region in the $MoS_2$ side of the $VSe_2/MoS_2$ interface is created, leading to the subsequent formation of a Schottky barrier. Application of an electric field or electric gating can promote this charge transfer process and hence alter interfacial magnetic coupling (or the interfacial exchange anisotropy) in the $VSe_2/MoS_2$ heterostructure. In other words, the interfacial exchange anisotropy and hence spin transport could be controlled by electric fields. A similar perspective can be expected for the $MnSe_2/MoSe_2$ heterostructure [171]. These 2D-TMD heterostructures may find potential applications in energy conversion and 2D vdW spintronics.

Unlike monolayer $VSe_2$ that mostly stabilizes in a 1T structure [40,95-97], $VS_2$ has been reported to favorably form in a 2H structure and to exhibit FM order at room temperature down to the 2D limit [172-175]. This 2D-TMD is therefore considered an intrinsic ferromagnetic semiconductor. First-principles calculations have revealed a significant ME effect in bilayer $VS_2$ [176]. The calculations show that the magnetic moments couple ferromagnetically within each $VS_2$ layer, but couple antiferromagnetically between the two layers. The coupling between ferroelectricity and antiferromagnetism via a ferrovalley has been predicted to enable an electronic field control of magnetism in the bilayer $VS_2$.

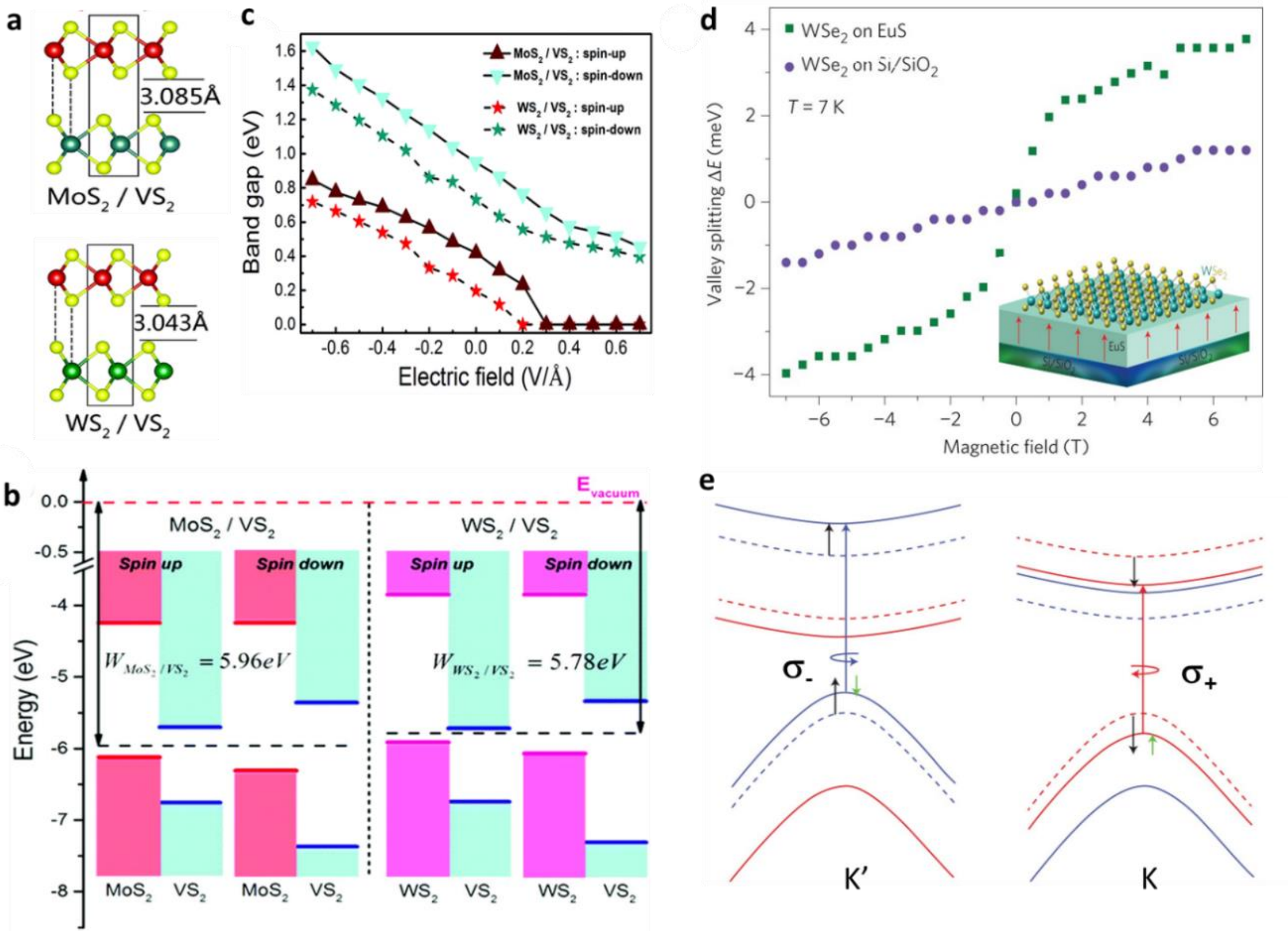

**Figure 10.** (**a**) The side view, (**b**) the band edges and work functions, and (**c**) the electric field-modulated majority and minority band gaps of 2D $MoS_2/VS_2$ and $WS_2/VS_2$ heterostructures. Panels (a,b,c) reproduced with permission from Ref. [177]. (**d**) When monolayer $WSe_2$ is placed on an EuS substrate, valley splitting in monolayer $WSe_2$ is greatly enhanced, with an initial slope of 2.5 meV $T^{-1}$ at $-1$ T $< \mu_0 H <$ +1 T. This is in contrast with the linear response of valley splitting to the external magnetic fields observed for the $WSe_2$ monolayer on Si/$SiO_2$ substrate. (**e**) A schematic energy diagram at the *K* and *K′* valleys showing energy splitting of the bands in the exchange coupled EuS/$WSe_2$ heterostructure and the corresponding A exciton transitions at the *K* and *K′* valleys. Panels (d,e) reproduced with permission from Ref. [178].

DFT calculations by Du *et al.* revealed that $MoS_2/VS_2$ and $WS_2/VS_2$ heterostructures exhibit semiconducting and ferromagnetic properties, with Curie temperatures exceeding 300 K, and their electric and magnetic functionalities can be modulated by electric fields [177]. By forming $MoS_2/VS_2$ and $WS_2/VS_2$ interfaces (Fig. 10a), charge transfer occurs across the interface from $MoS_2$ or $WS_2$ to $VS_2$ since both $MoS_2$ and $WS_2$ have smaller work functions compared to $VS_2$. Electrons are accumulated in the $VS_2$ layer, while holes are occupied in the $MoS_2$ or $WS_2$ layer. The authors show that the application of a positive external electric field can promote a semiconductor to a half-metal transition. Depending upon applied negative and positive external electric fields, type-II to type-I and type-II to type-III band alignment transitions can appear. The bandgap is largely tunable with varying negative external electric fields (Fig. 10b,c). Experiments are needed to validate these predictions.

For valleytronics applications, the use of an interfacial magnetic exchange field (IMEF) from a ferromagnetic vdW material (like EuS) has been shown to greatly enhance valley splitting in monolayer $WSe_2$ or $WS_2$ [142,178]. Figure 10d shows the enhanced valley splitting $\Delta E$ in the $EuS/WSe_2$ heterostructure relative to the monolayer $WSe_2$ on $Si/SiO_2$ substrate [178]. Due to the MPE induced by the EuS layer, energy level shifts occur, which can be interpreted based on the energy diagram at the K and K′ valleys (Fig. 10e). Under an external magnetic field, the spin, atomic orbital, and valley orbital magnetic moments were found to give rise to these energy shifts. However, it is worth noticing that the large $\Delta E$ in the $EuS/WSe_2$ heterostructure can be achieved only at low temperatures in the FM region of EuS (below its Curie temperature, ~50 K). In this regard, using the IMEF of monolayer $VSe_2$ [40] or $MnSe_2$ [42] in $VSe_2/WSe_2$ ($WS_2$) or $MnSe_2/WSe_2$ ($WS_2$) heterostructures to tune valley splitting in monolayer $WSe_2$ or $WS_2$ would be more promising because the magnetic field needed to flip the magnetic moments of monolayer $VSe_2$ or $MnSe_2$ (~ 500 Oe) is about 20 times smaller than that of EuS (~10 kOe), which can be achieved at room temperature unlike the case of $EuS/WSe_2$ ($WS_2$) where EuS is ferromagnetic only below 50 K. First principles calculations have predicted a quantum spin Hall effect in TMDs, highlighting the prospective use of these TMDs and 2D dielectric layers to create a topological field effect transistors that can be rapidly switched off by an electric field [179]. In this context, the observation of both the exchange bias and magnetic proximity effects, as well as the light-controlled ferromagnetism in the $VSe_2/MoS_2$ heterostructures [35,40,180], points to a novel electrically and optically tunable interfacial magnetic exchange coupling in these TMD heterostructures through which to control valleytronic responses of the $MoS_2$ monolayer.

It is noteworthy that depending on how a TMD monolayer is stacked on a magnetic substrate, the valley splitting effect varies considerably. Based on DFT calculations, Li *et al.* demonstrated stacking-dependent valley splitting $\Delta E$ in 2D $MoTe_2/MnSe_2$ heterostructures [171]. The $\Delta E$ value of the monolayer $MoTe_2$ can reach 106 meV when the monolayer $MoTe_2$ is optimally stacked on the monolayer $MnSe_2$, which has been reported to exhibit FM order at room temperature [42]. This value of $\Delta E$ is equivalent to the effective Zeeman splitting caused by an external magnetic field as large as 530 T. Interestingly, the application of external vertical-stress and biaxial compressive strain on monolayer $MnSe_2$ has been shown to enhance $\Delta E$ of the monolayer $MoTe_2$. On the other hand, the application of a biaxial tensile strain increased the Curie temperature of the monolayer $MnSe_2$ from 266 K to 353 K, while retaining a large $\Delta E$ of 72 meV. Like $VSe_2/MoS_2$ and $VS_2/MoS_2$ heterostructures, charge transfer may occur across the $MnSe_2/MoTe_2$ interface and mediate its interfacial property by electric means through a voltage-modulated directional charge transfer mechanism. All these interesting predictions need to be verified by experiments.

## 3. Opportunities and Challenges

Table 1 was built to list potential material candidates to exploit 2D vdW magnets and heterostructures with electrically tunable magnetic functionalities for applications in spintronics, spin-caloritronics, and valleytronics. From a materials engineering perspective, emerging opportunities (for both fundamental science and applications) and technical challenges are discussed, as well as possible approaches to addressing these challenges. It is worth noting in Table 1 that the magnetic ordering temperature ($T_C$ or $T_N$) of a vdW material (also in its heterostructures) determines its application's temperature range. Therefore, we highlight the magnetic ordering temperatures of several theoretically predicted and/or experimentally realized electrically tunable vdW magnetic systems ranging from metals to semiconductors and heterostructures. We evaluate the electrically tunable magnetic functionalities, novel emerging effects (exchange bias - EB, or magnetoelectric - ME, or magnetic skyrmions), and the air stable properties of these materials with a focus on their perspective applications in spintronics, caloritronics, and valleytronics.

**Table 1.** Ordering temperatures and properties of electrically tunable 2D vdW magnets and heterostructures (discovered experimentally or theoretically, or both).

| Materials | Ordering/operating temperatures | Remarks | References |
|---|---|---|---|
| **Metals** | | | |
| $Fe_3GeTe_2$ (NF, SC) | $T < T_C$ ~130 - 230 K | Gate tunability in $M_S$, $H_C$, and $T_C$; Sensitive to air. | [87,88] |
| $Fe_5GeTe_2$ (NF, SC) | $T < T_C$ ~300 - 350 K | Gate tunability in $M_S$, $H_C$, and $T_C$; Less sensitive to air. | [83] |
| $VSe_2$ (ML, BL) | $T < T_C$ ~270 - 350 K | Unexplored ECM experiments; Sensitive to air. | [40,105] |
| $MnSe_2$ (ML, BL) | $T < T_C$ ~266 - 350 K | Unexplored ECM experiments; Sensitive to air. | [42,181] |
| $CrSe_2$ (ML) | $T < T_C$ ~300 K | Unexplored ECM experiments; Stable in air. | [45,106] |
| $FeSe_2$ (ML) | $T < T_C$ ~300 K | Unexplored ECM experiments; Stable in air. | [106] |
| MnSSe | $T < T_C$ ~72 K | Unexplored ECM experiments; Stable in air. | [182] |
| $Fe_{1/3}NbS_2$ (SC) | $T < T_N$ ~42 K | Electric control of magnetic state at low temperatures; Stable to air. | [108] |
| **Semiconductors** | | | |
| $CrI_3$ (BL) | $T < T_N$ ~45 K | Magnetic switching by gate voltages; Large ME effect; Sensitive to air. | [110-113] |
| Twisted $CrI_3$ (BL) | $T < T_N$ ~45 K | Magnetic switching by gate voltages; Enhanced ME effect; Sensitive to air. | [114,115] |
| $CrI_3$ (ML) | $T < T_C$ ~45 K | Unexplored ECM experiments; Sensitive to air. | [113] |
| $CrBr_3$ | $T < T_C$ ~34 K | Unexplored ECM experiments; Sensitive to air. | [43,183] |
| $CrCl_3$ | $T < T_N$ ~20 K | Unexplored ECM experiments; Sensitive to air. | [183] |
| $Cr_2Ge_2Te_6$ (BL, SC) | $T < T_C$ ~30-68 K | Gate tunability in $M_S$, $H_C$, and $T_C$; Sensitive to air. | [118-120] |
| V-doped TMDs ($V$-$WSe_2$, $V$-$WS_2$) | $T < T_C$ ~300-350 K | Gate or light tunability in $M_S$, $H_C$, and $T_C$; Less sensitive to air. | [50-55] |
| Co-doped ZnO (ML) | $T < T_C$ ~300-350 K | Predicted Theory; Unexplored ECM experiments; Stable in air. | [124,125] |
| **Heterostructures** | | | |
| $Fe_5GeTe_2/FePS_3$ | $T < T_N$ ~30-60 K | Gate tunability in $M_S$, $H_C$, and $T_C$; EB effect; Less sensitive to air. | [132] |
| $Fe_3GeTe_2/FePS_3$ | $T < T_N$ ~110 K | Unexplored ECM experiments; EB effect; Sensitive to air. | [184] |
| $Fe_3GeTe_2/MnPS_3$ | $T < T_N$ ~78 K | Gate tunability in $M_S$, $H_C$, and $T_C$; EB effect; Sensitive to air. | [128] |
| $Fe_3GeTe_2/CrCl_3$ (30 nm/15 nm) | $T < T_N$ ~14 K | Unexplored ECM experiments; EB effect; Sensitive to air. | [126] |
| CrTe/MnTe (5 nm/40 nm) | $T < T_N$ ~70 K | Unexplored ECM experiments; EB effect; Stable to air. | [185] |
| $Fe_3GeTe_2/Cr_2Ge_2Te_6$ | $T < T_C$ ~30-68 K | Experimentally unexplored electric control of skyrmion multi-states; Charge transfer; Sensitive to air. | [79] |
| $Fe_3GeTe_2/\alpha$-$In_2Se_3$ | $T < T_C$ ~100-230 K; $T_F > 600$ K | Experimentally unexplored ferroelectric control of skyrmion states; ME effect; Charge transfer; Sensitive to air. | [136] |
| $Fe_3GeTe_2/WSe_2$ | $T < T_C$ ~100-230 K | Experimentally unexplored electric control of valleytronic states; Charge transfer; Sensitive to air. | [147] |
| Gr/ML $CrI_3$/Gr | $T < T_C$ ~45 K | Gate tunability in $M_S$, $H_C$, and $T_C$; Less sensitive to air. | [149] |
| Gr/BL $CrI_3$/Gr | $T < T_N$ ~45 K | Magnetic switching by gate voltage; Less sensitive to air. | [150] |
| BL $CrI_3$/$WSe_2$ | $T < T_N$ ~45 K | Electric control of antiferromagnetic resonance; Experimentally unexplored electric control of valleytronic states; Charge transfer; Sensitive to air. | [143,146,151,152] |
| $CrI_3$/$WSe_2$/$CrI_3$ | $T < T_C$ ~45 K | Experimentally unexplored electric control of valleytronic states; Charge transfer; Sensitive to air. | [153] |
| Twisted $CrI_3$/$WSe_2$ | $T < T_C$ ~45 K | Enhanced electric control of valleytronic states; Charge transfer; Sensitive to air. | [154] |
| $CrCl_3$/$WTe_2$ | $T < T_N$ ~20 K | Experimentally unexplored electric control of skyrmion states; Charge transfer; Sensitive to air. | [158] |
| ML $CrI_3$/$\alpha$-$In_2Se_3$ | $T < T_C$ ~45 K | Experimentally unexplored ferroelectric control of magnetism; Charge transfer; Sensitive to air. | [155] |
| ML $CrI_3/Sc_2CO_2$ | $T < T_C$ ~45 K | Experimentally unexplored ferroelectric control of magnetism; Charge transfer; Sensitive to air. | [157] |
| $CrCl_3/CuInP_2S_6$ | $T < T_N$ ~20 K | Unexplored ECM experiments; Charge transfer; Sensitive to air. | [158] |
| $MnPS_3/CuInP_2S_6$ | $T < T_N$ ~80 K | Unexplored ECM experiments; Charge transfer; Sensitive to air. | [159] |
| $Cr_2Ge_2Te_6/\alpha\text{-}In_2Se_3$ | $T < T_C$ ~30-68 K | Experimentally unexplored ferroelectric control of magnetism; Charge transfer; Sensitive to air. | [161] |
| $Cr_2Ge_2Te_6/Sc_2CO_2$ | $T < T_C$ ~30-68 K | Experimentally unexplored ferroelectric control of magnetism; Charge transfer; Sensitive to air. | [162] |
| $Cr_2Ge_2Te_6/BiSbTeSe_2$ | $T < T_C$ ~30-68 K | Experimentally unexplored ferroelectric control of magnetism; Charge transfer; Sensitive to air. | [163] |
| $Cr_2Ge_2Te_6/MoSe_2$ | $T < T_C$ ~30-68 K | Electric/magnetic control of valleytronic states; Charge transfer; Less sensitive to air. | [164] |
| $Cr_2Ge_2Te_6/WSe_2$ | $T < T_C$ ~30-68 K | Unexplored electric/magnetic control of valleytronic states; Charge transfer; Less sensitive to air. | [165] |
| $VSe_2/MoTe_2$ | $T < T_C$ ~300-350 K | Experimentally unexplored electric control of valleytronic states; Charge transfer; Less sensitive to air. | [168] |
| $VSe_2/MoS_2$ | $T < T_C$ ~300-350 K | Experimentally unexplored electric control of valleytronic states; EB effect; Charge transfer; Less sensitive to air. | [40,180] |
| 2H-$VSe_2/Sc_2CO_2$ | $T < T_C$ ~300 K | Experimentally unexplored ferroelectric control of valleytronic states; Charge transfer; Less sensitive to air. | [188,189] |
| $MnSe_2/MoSe_2$ | $T < T_C$ ~300 K | Experimentally unexplored electric control of valleytronic states; Charge transfer; Less sensitive to air. | [45,171] |
| $VS_2/MoS_2$ or $WS_2$ | $T < T_C$ ~300-350 K | Experimentally unexplored electric control of valleytronic | [176] |
| | | states; Charge transfer; Less sensitive to air. | |
| $CrSnSe_3/WSe_2$ | $T < T_C$ ~60 K | Experimentally unexplored electric control of valleytronic states; Charge transfer; Less sensitive to air. | [186,187] |
| α-$RuCl_3/CuInP_2S_6$ | $T < T_C$ ~89 K | Experimentally unexplored ferroelectric polarization control of magnetism; Charge transfer; Stable to air. | [190] |
| α-$In_2Se_3$/X (X = Co, Fe) | $T < T_C$ ~89 K | Experimentally unexplored Control of magnetic anisotropy by ferroelectric polarization; Charge transfer; Stable to air. | [191] |

**Note:** *ECM-Electric control of magnetism; ML-Monolayer; BL-Bilayer;* $T_C$*-Curie temperature;* $T_N$*-Néel temperature.*

### *3.1. Spintronics*

Charge-based transistors, which are primarily based on silicon materials, are employed to generate electric switches and are the key components in integrated circuits. Unlike these charge-based transistors, transistors based on spin, called "spin transistors," may offer non-volatile data storage and higher energy efficiency [18,22,150,192,193]. A wide range of 2D vdW magnets and heterostructures have been proposed and explored for the fabrication of spin transistors [18-24,150,192-94].

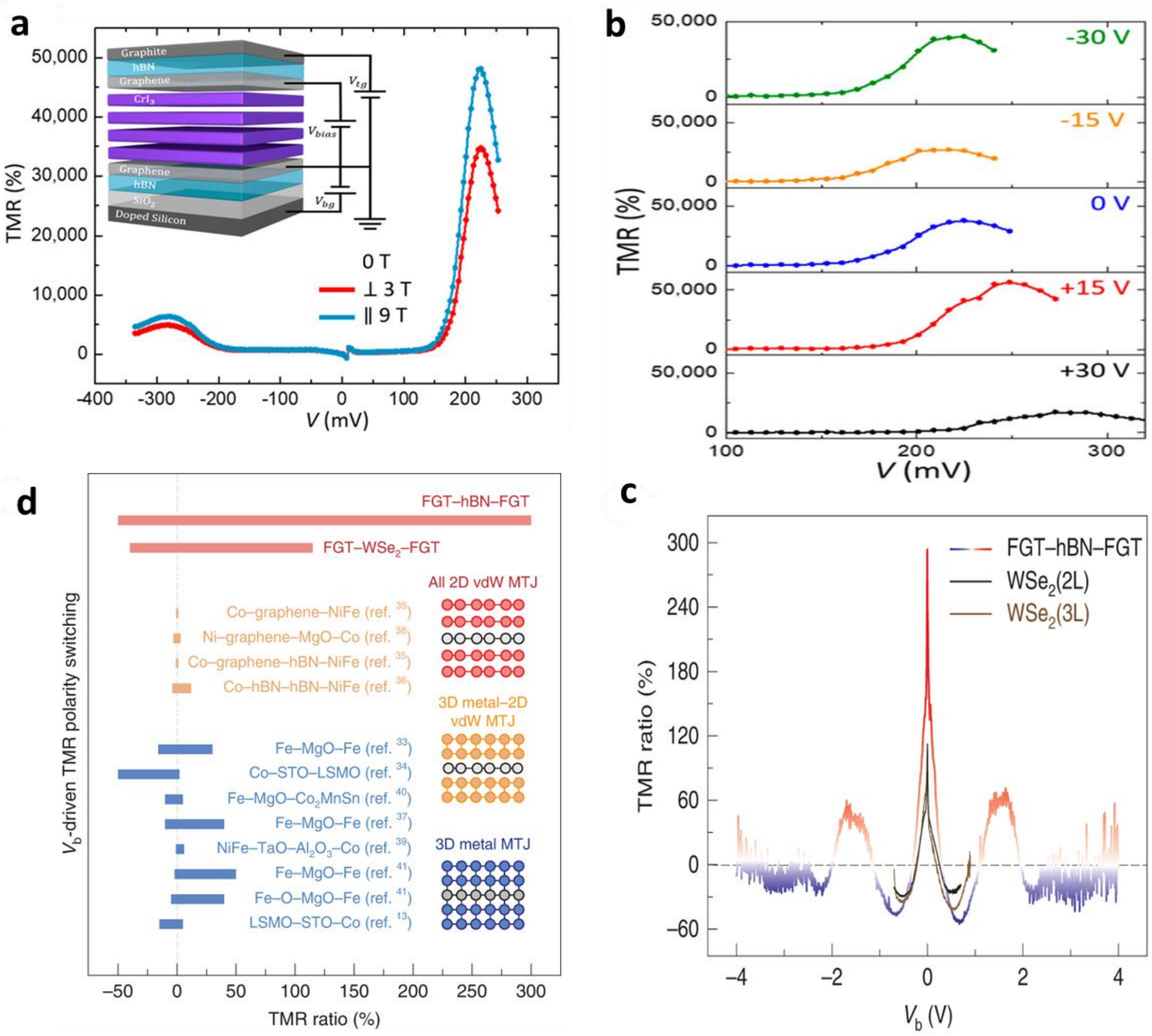

**Figure 11.** (**a**) TMR ratio as a function of *V* derived from the $I_t$-*V* data; inset shows a schematic of a four-layer $CrI_3$ sf-MTJ device including two monolayer graphene contacts and top/bottom

gates; (**b**) TMR ratio as a function of *V* at a series of $V_{bg}$; panels (a,b) reproduced with permission from Ref. [150]. (**c**) The $V_b$-dependent TMR ratio variations for the FGT-hBN-FGT device along with the FGT-$WSe_2$-FGT devices with bilayer and trilayer $WSe_2$ as a tunnel barrier. (**d**) Bias-dependent TMR polarity switching in various MTJs. Panels (c,d) reproduced with permission from Ref. [197].

Electrically tunable spin-based devices or magnetic tunnel junctions (TMJs) utilizing graphite/BL-$CrI_3$/graphite heterojunctions have been designed, fabricated, and tested [195]. In the graphite/BL-$CrI_3$/graphite heterostructure, the bilayer $CrI_3$ acts as a spin filter tunnel barrier so that application of an appropriate magnetic field can switch the magnetic order from the AFM to FM state, resulting in a large change in resistance, which is known as giant tunneling magnetoresistance (TMR). In its initial AFM state (in zero applied magnetic field), bilayer $CrI_3$ hinders tunneling of electrons through adjacent layers, leading to the high resistance state. A sufficiently high magnetic field applied to bilayer $CrI_3$ will flip spins from the AFM to the FM state, allowing large number of electrons to tunnel through the adjacent layers, reducing the resistance of the heterojunction/device. This TMR can be electrically controlled, reaching a maximum ratio of 530% at 290 mV bias voltage [195]. The extremely large values of TMR, 3,200% and 19,000%, have been reported for devices using trilayer and four-layer $CrI_3$, respectively. Using a dual-gated graphene/four-layer $CrI_3$/graphene device [150], the magnetic state of the $CrI_3$ layer and hence the TMR of the device could also be controlled electronically. The extremely large values of TMR (> 45,000%) have been reported for these MTJs (Fig. 11a,b). Recently, DFT calculations performed by Zhu *et al.* have shown a large TMR ratio of 2,420% in a graphene/bilayer $CoBr_2$/graphene-based device at zero bias [196]. The application of a small bias voltage (~0.2 V) can increase the TMR ratio up to 38,000%. However, this prediction needs to be validated by experiments. While the 2D vdW intrinsic ferromagnets $CrI_3$ and $Cr_2Ge_2Te_6$ show excellent gate-modulated magnetic and TMR properties, these materials and their heterostructures are restricted to operate at low temperatures (well below 100 K) due to their low magnetic ordering temperatures (see Table 1). In addition, these materials are extremely air sensitive, and an extra care is needed to handle them while fabricating and testing their devices. These systems are therefore not desirable for practical applications in spintronics and other spin-based devices. Nonetheless, they are the excellent model systems for exploring and harnessing new functionalities and spin transport phenomena down to the 2D limit, which will facilitate the development of novel multifunctional nanomaterials and nanodevices, as well as expand their applications.

Although $Fe_3GeTe_2$ and $Fe_5GeTe_2$ are metallic in nature, these 2D vdW intrinsic ferromagnets show their strong magnetic responses to external stimuli (electric field, light, and strain) [39,83,86,88,133]. These materials and their heterostructures have been widely researched during the last few years [20-24,29-31,132]. In a recent study, Min *et al.* have demonstrated electrically tunable, highly transparent spin injection and detection across $Fe_3GeTe_2/WSe_2/Fe_3GeTe_2$ and $Fe_3GeTe_2/hBN/Fe_3GeTe_2$ interfaces [197]. Applying an appropriate bias voltage can modulate the net spin polarization of the injected carriers and reverse its polarity, leading to a sign change of the TMR (Fig. 11c). This effect has been attributed to the contribution of high-energy localized spin states in $Fe_3GeTe_2$. As compared to other non-vdW material-based MTJs, the $Fe_3GeTe_2/WSe_2/Fe_3GeTe_2$ and $Fe_3GeTe_2/hBN/Fe_3GeTe_2$-based MTJ devices have shown the widest variation in TMR ratio with clear sign changes (Fig. 11d), emphasizing the importance of atomically sharp interfaces in 2D vdW magnets.

In spin-based device applications, generating a spin-polarized current within a ferromagnetic material and retaining spin polarization across its interface with a non-magnetic material with strong spin orbit coupling (Pt, W) are essential. Spin orbit torque (SOT), which is based on a spin current to manipulate the magnetization of a magnetic material, has been widely explored in spintronic devices [30,31]. The origin of SOT comes from the spin Hall effect (SHE) and Rashba effect, which can generate spin currents from a sourcing charge current. To achieve magnetization switching of a ferromagnetic material, ferromagnet/heavy metal (FM/HM) bilayer systems are often required to break inversion symmetry. In SOT-based devices, the quality of such FM/HM interfaces is crucial [30]. Owing to atomically flat vdW interfaces, highly efficient SOT and magnetization switching in $Fe_3GeTe_2$, $CrI_3$, and $Cr_2Ge_2Te_6$ have been experimentally realized [198-200]. Dolui *et al.* employed a $TaSe_2/CrI_3$ vdW heterostructure to switch from the AFM to FM state based on the SOT mechanism [201]. Utilizing $WTe_2$ and FGT with low symmetry to form a $WTe_2$/FGT vdW heterostructure can also yield a deterministic switching of the magnetization [202]. The SOT efficiency reported for $Fe_3GeTe_2$/Pt is greater than that of heterostructures containing 3D ferrimagnetic insulators and comparable to that of the best heterostructures containing 3D ferromagnetic metals [203]. It has been theoretically shown that the directional charge transfer from a metallic vdW magnet (e.g., $Fe_3GeTe_2$) to a semiconducting vdW magnet (e.g., $Cr_2Ge_2Te_6$) can alter the magnetic anisotropy energy and hence the coercive field of the metal (e.g., $Fe_3GeTe_2$), suggesting a new approach for electrically manipulating the magnetic properties of ferromagnetic electrodes (e.g., $Fe_3GeTe_2$) in vdW magnetic heterojunctions through the voltage-controllable charge transfer mechanism [133].

It is anticipated that better electric control and thermal perturbation of 2D vdW heterostructures enables magnetization switching of the FM layer with the use of lower currents as compared to non-vdW material-based junctions in SOT devices. While the surface magnetic properties of FM electrodes like $Fe_3GeTe_2$ are crucial for achieving large SV and SOT effects, these 2D vdW magnets are sensitive to air and their surface magnetic properties become degraded rapidly when oxidized, thus reducing the overall performance of FGT-based devices. So, careful attention must be paid to preparing these 2D vdW materials and their heterostructures, and for making the corresponding devices.

Magnetically-doped TMD monolayers (like V-doped $WSe_2$, V-doped $WS_2$, and V-doped $MoSe_2$) are emerging as a novel class of 2D-DMSs that order ferromagnetically at room temperature [50-53,204]. Their electrically and optically tunable magnetic properties make them excellent candidates for use in 2D vdW transistors operating at room temperature [19,23,52,54,205-209]. In a recent study, Ghiasi *et al.* demonstrated charge-to-spin conversion across a monolayer $WS_2$/graphene interface due to the Rashba-Edelstein effect (REE) and that the current-induced spin polarization can be controlled by gate voltages [210]. In this context, the use of 2D-DMSs based on V-doped $WS_2$ monolayers or V-doped $WSe_2$ monolayers may be of great interest as they have been predicted to boost spin-charge conversion efficiency [211,212]. Furthermore, it is possible to manipulate the spin-charge conversion process by optical means, as these monolayers have shown optically tunable magnetic properties [54]. However, the saturation magnetizations of 2D DMSs are still limited to approximately $10^{-5}$ emu $cm^{-2}$, thus hampering their practical use. While FM properties can be enhanced and optimized by varying concentration of magnetic dopants, heavy doping has been reported to strongly suppress FM order and photoluminescence [50,52]. Further efforts are therefore needed to address: (i) the uncertain origin of room temperature ferromagnetism, i.e., coexistence of intrinsic and extrinsic (defects-induced) FM, due to a lack of structural analysis of atomic defects; (ii) solubility limit to only a few percent without forming aggregation; and (iii) activation of short-range antiferromagnetic (AFM) coupling in the high doping regime, which negates the improvement of the magnetic moment.

The use of ferroelectric polarization of a ferroelectric vdW material (e.g., $\alpha$-$In_2Se_3$, $Sc_2CO_2$) to control/switch the magnetization of a 2D vdW magnet (e.g., $Fe_3GeTe_2$, $CrI_3$, and $Cr_2Ge_2Te_6$) is also a very promising approach for spintronic device applications [136,155-157,161,162]. The electric control of magnetic skyrmions in the 2D vdW ferromagnet $Fe_3GeTe_2$ using the ferroelectric polarization of a 2D vdW ferroelectric material like $\alpha$-$In_2Se_3$ has indeed opened a new opportunity for applications in low-power memory and logic devices [136]. However, most of the works reported to date have been based on DFT calculations and modelling [155-157,161,162]. Solid evidence is lacking, and experimental studies are therefore needed to validate these predictions. Stacking ferroelectric 2D vdW materials with 2D vdW magnets to form novel 2D magnetoelectric heterostructures is an important but challenging task. Effects of intercalation, interdiffusion, twisting, moiré patterns, etc., may occur in these heterostructures, altering the interfacial magnetic properties and the magnetoelectric coupling [213-216]. Other effects such as magnetic proximity and charge transfer are also important and must be studied thoroughly [213,216]. Having examined the electrically tunable magnetic properties of the reported 2D vdW materials and heterostructures, we outline several new and exciting opportunities in the field of 2D vdW spintronics, as well as the technical challenges that warrant further studies (Scheme 1).

**Scheme 1.** Opportunities and challenges in the field of 2D vdW spintronics.

<table>
<tr><th colspan="2">Spintronics</th></tr>
<tr><th>Opportunities</th><th>Challenges</th></tr>
<tr>
<td>

- Spin transistors based on 2D room-temperature ferromagnetic TMD semiconductors (e.g., V-doped $WSe_2$)

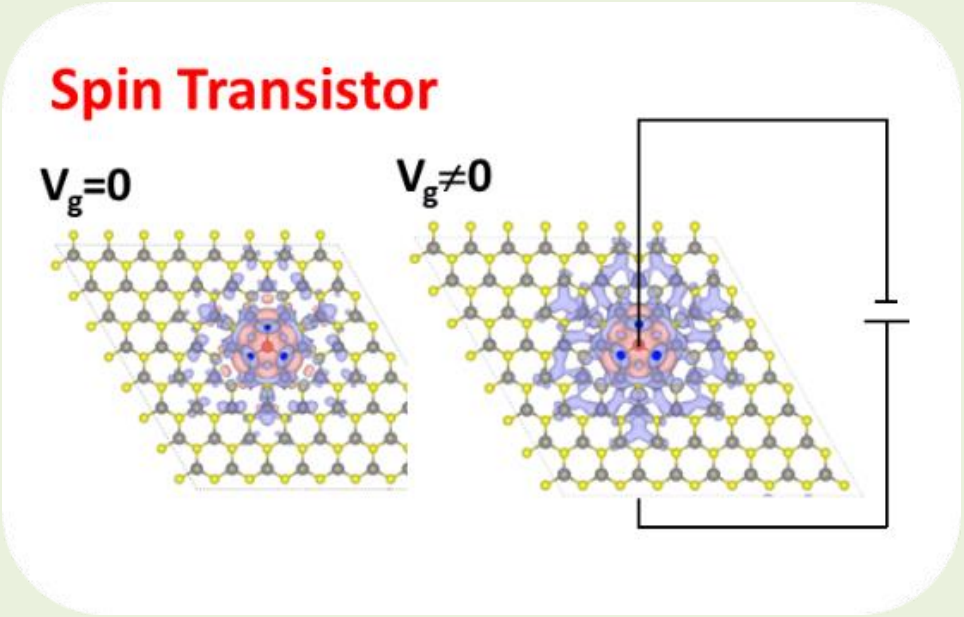

- Electrically and optically tunable magnetism in diluted magnetic 2D-TMDs for sensing and spin-logic applications

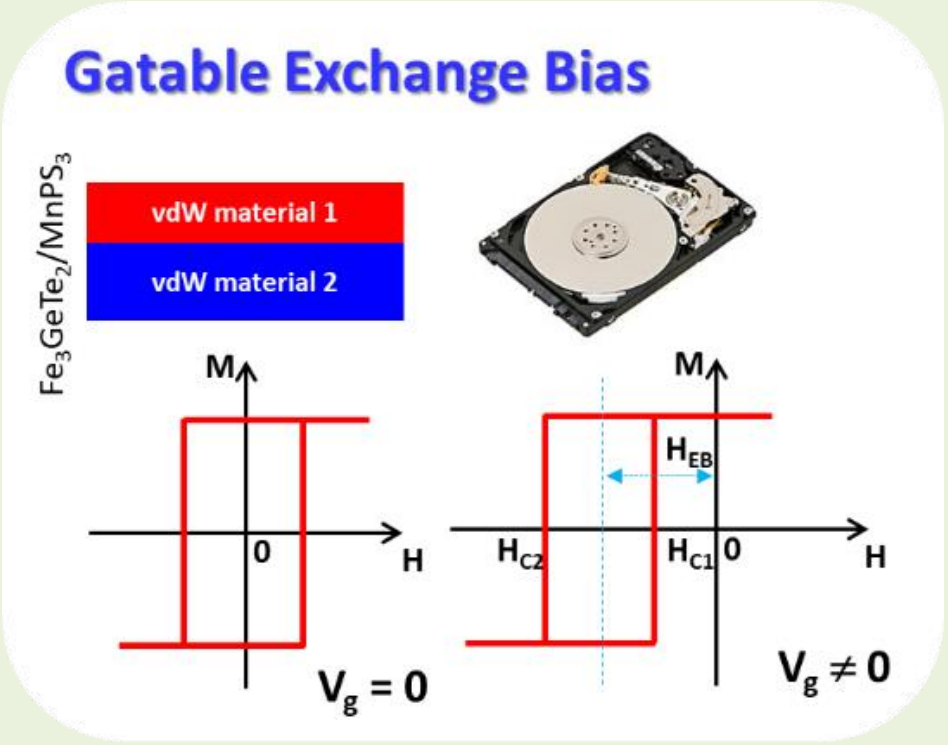

- Electrically tunable interfacial magnetism (exchange bias or exchange anisotropy) in vdW FM/vdW AFM heterostructures (e.g.,

</td>
<td>

- Weak ferromagnetism
- Difficulty controlling magnetic dopants/defects in TMDs to achieve reproducible magnetic properties (e.g., $M_S$, $H_C$)
- Strong suppression of ferromagnetism and photoluminescence in heavily doped 2D-TMDs
- Air instability (magnetic signals are degradable when exposed to air)
- Difficulty with integration and fabrication of 2D-TMD based devices
- Works at low temperatures well below 300 K, depend on stacking quality and thickness, self-intercalation, air instability (magnetic degradation), fabrication difficulty
- Weak interfacial coupling between layers
- Need to seek vdW materials that order magnetically at temperatures ($T_C$, $T_N$) around room temperature

</td>
</tr>
</table>

| | |
|---|---|
| $Fe_3GeTe_2/FePS_3$)<br>• 2D vdW magnetic materials for spin orbit torque (SOT) based spintronics (e.g., $Fe_3GeTe_2/Pt$)<br>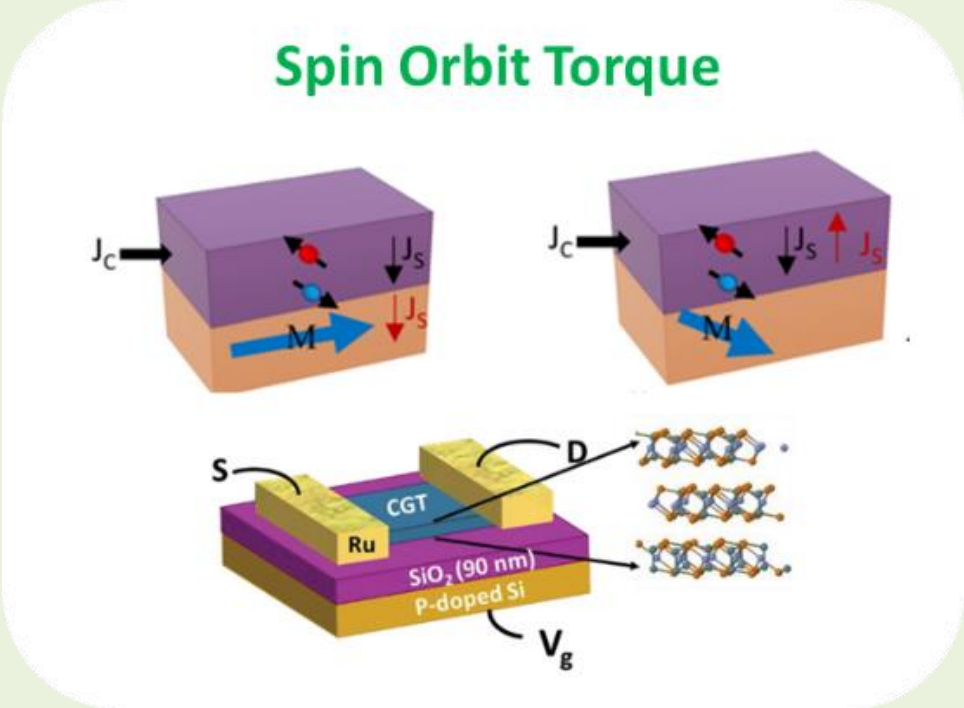<br>• Ferroelectric control of 2D vdW magnetism (e.g., $Cr_2Ge_2Te_6/\alpha\text{-}In_2Se_3$, $MnPS_3/CuInP_2S_6$)<br>• Electric control of skyrmion states across vdW magnetic interfaces (e.g., $Fe_3GeTe_2/Cr_2Ge_2Te_6$)<br>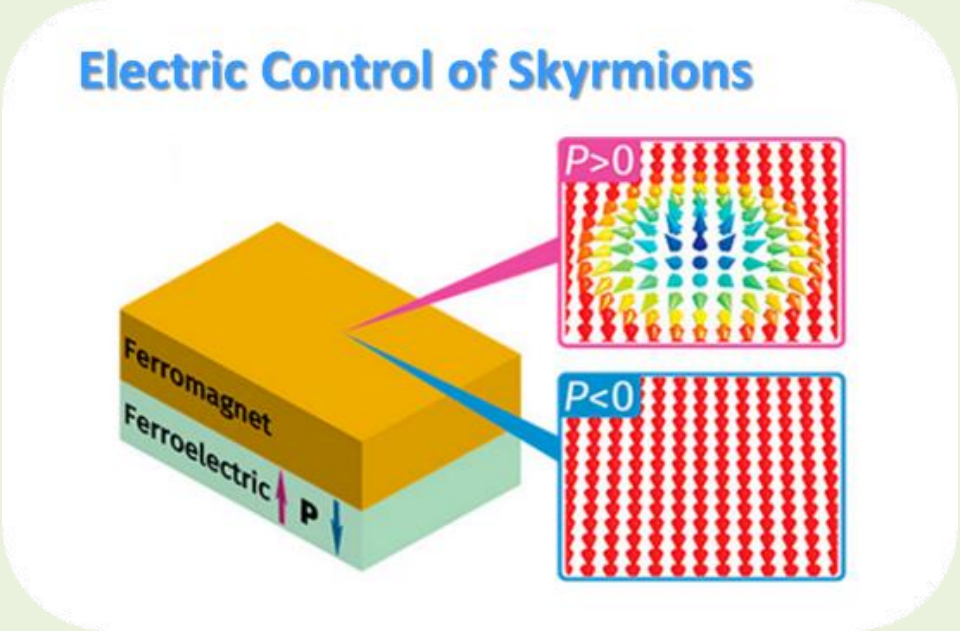<br>• Ferroelectric control of skyrmion states across vdW magnetic/vdW ferroelectric interfaces (e.g., $Fe_3GeTe_2/\alpha\text{-}In_2Se_3$) | • Difficulty synthesizing and stacking 2D vdW magnetic and ferroelectric materials together in heterostructures<br>• Need to achieve low-power read-out of SOT devices; interface-dependent SOT effects; difficulty engineering atomically flat vdW/non-vdW interfaces<br>• Interatomic diffusion at vdW magnet/metal interfaces, interface oxidization (magnetic degradation), short lifetime of SOT-based devices<br>• Complicated effects of charge transfer and magnetic proximity across heterostructure interfaces<br>• Undesired effects of intercalation, interdiffusion, twisting, moiré patterns, etc.<br>• Formation of unstable magnetic skyrmions; difficulty controlling size and shape of magnetic skyrmions, as well as large scale device fabrication/integration<br>• Large discrepancies between theoretical predictions and experiments |

**Note:** *Image #4 is reproduced from Ref. [136].*

### *3.2. Spin-Caloritronics*

The spin Seebeck effect (SSE) is the generation of a thermally driven spin current in a magnetic material subject to an applied magnetic field [217,218]. It has potential applications in thermo-electric generation, spin-detection, and temperature sensing [218]. The SSE is often assessed by the inverse spin Hall effect (ISHE), which converts spin current into an electric field by means of spin-orbit scattering (Fig. 12a). By studying the ISHE, we can indirectly determine the spin voltage developed due to the SSE. Platinum (Pt) is an efficient ISHE material and frequently used as electrodes. Owing to its ferrimagnetic insulating characteristics and low-damping coefficient, $Y_3Fe_5O_{12}$ (YIG) has been extensively explored as a benchmark material for generating a pure spin current via the SSE and for SSE-related studies [219-221]. However, the large conductivity mismatch between the insulator (YIG) and metal (Pt) hampers the spin-to-charge conversion efficiency and hence the SSE [218]. To resolve this, a thin semiconducting layer with long spin diffusion length like $C_{60}$ can be placed between YIG and Pt, as demonstrated by Kalappattil *et al.*, to reduce the conductivity mismatch and give rise to an enhanced SSE (an increase of 600%) [221]. Following this approach, Lee *et al.* [222] showed that the insertion of a semiconducting TMD monolayer such as $WSe_2$ in between YIG and Pt increases the SSE in the YIG/monolayer $WSe_2$/Pt heterostructure significantly (up to 500%) as a result of the reduced conductivity mismatch. The MPE of YIG on $WSe_2$ monolayer magnetism has been theoretically and experimentally shown to impact spin transport and hence the spin-to-charge conversion efficiency in YIG/ML-$WSe_2$/Pt [211,223]. DFT calculations have shown that the presence of the interlayer $WSe_2$ increases the overall total density of states at the Fermi level (by 23%) in Fe/ML-$WSe_2$/Pt (Fig. 12b) relative to Fe/Pt (for simplicity, Fe was considered instead of YIG in these DFT calculations) [211]. The calculated spin Seebeck coefficient ($S_{spin}$) of Fe/TMD/Pt was found to be much greater compared to Fe/Pt. These findings suggest that the insertion of a semiconducting TMD monolayer in a FM/HM bilayer system not only reduces the conductivity mismatch, but also enhances the spin mixing conductance and hence the spin-to-charge conversion efficiency via the SSE. In this context, including a dilute magnetic 2D-TMD semiconductor such as V-doped $WS_2$ or V-doped $WSe_2$ monolayer between YIG and Pt has been predicted to enhance the spin mixing conductance and hence the SSE (see Fig. 12c) [211]. Since the magnetization of this magnetic TMD monolayer can be optically modulated [54], the SSE in such YIG/ML V-$WSe_2$/Pt heterostructures can also be controlled by optical means, opening a new research field called "Opto-Spin-Caloritronics" [34].

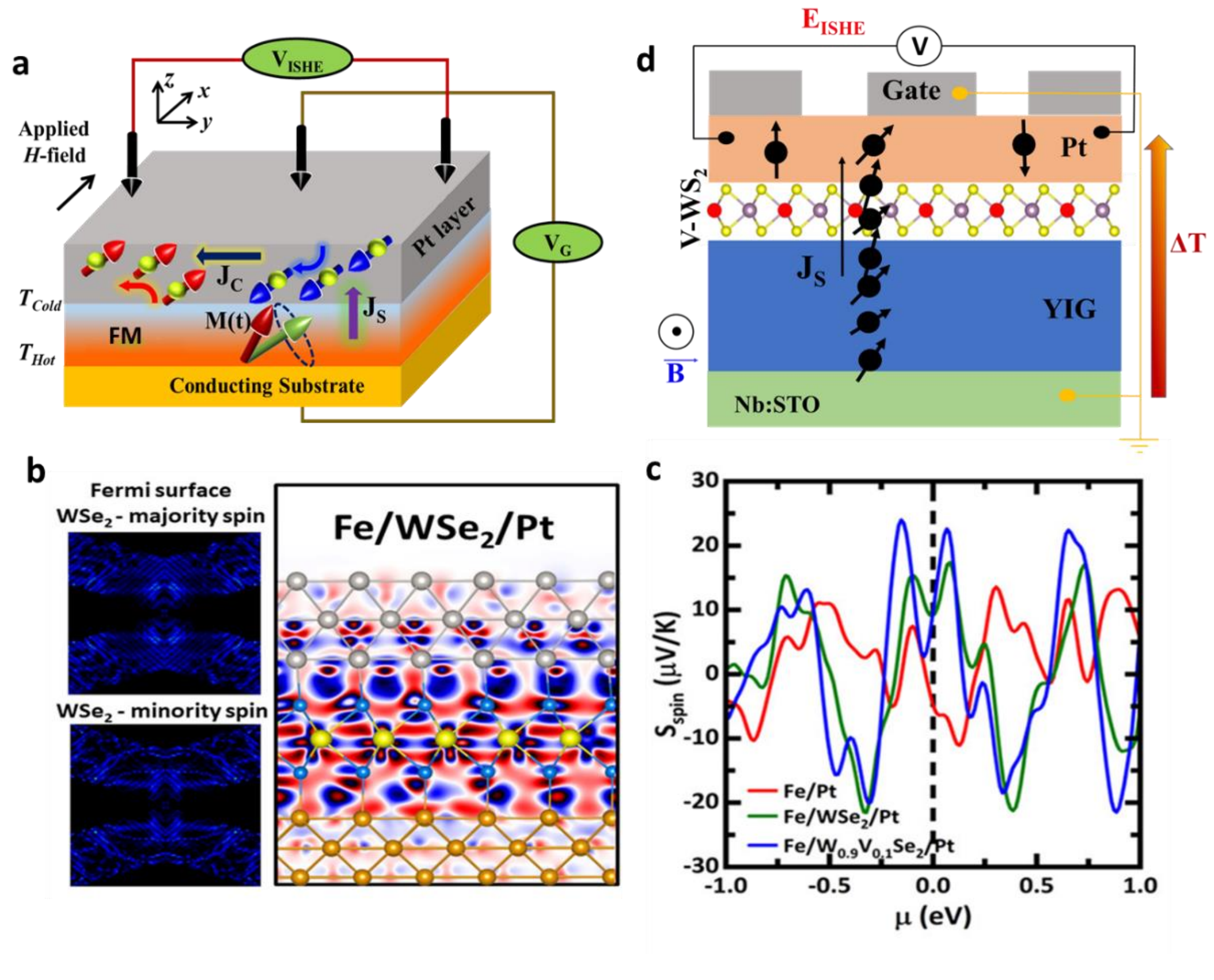

**Figure 12.** (**a**) The structure of a gating SSE device based on the YIG/Pt system; (**b**) Fermi surface state of Fe/$WSe_2$/Pt; (**c**) Spin Seebeck coefficient ($S_{spin}$) as a function of chemical potential μ for Fe/Pt, Fe/$WSe_2$/Pt, and Fe/V-$WSe_2$/Pt. Panels (b,c) reproduced with permission from Ref. [211]. (**d**) The structure of a proposed SSE gating device based on YIG/magnetic 2D-TMD/Pt

heterostructures in which the magnetization of the magnetic 2D-TMD can be controlled by gate voltage, enabling a gate-tunable spincaloritronic device.

Gate voltage control of magnetic functionality (e.g., voltage-controlled magnetic anisotropy) and hence the performance of magnetic devices based on anomalous Hall effect (AHE) and giant magnetoresistance (GMR) has been extensively explored over the last few years [35,224,225]. However, gate voltage control of thermally driven spin current and hence the spin-charge conversion efficiency through the SSE has been investigated much less [226-229]. Of these few studies, a recent one has shown the exciting possibility of using gate voltage to control magnetic anisotropy and hence the SSE sensitivity in PMNT/MgO/YIG/Pt systems [226]. Due to modification of the magnetic anisotropy of the YIG layer by electrostrain, low magnetic field sensitivity in SSE has been achieved. Since the magnetization of magnetic TMD monolayers is controllable by electric gating, as demonstrated by Duong *et al.* for V-doped $WSe_2$ monolayers (Fig. 7d-f) [52,55], it is possible to employ gate voltage to manipulate and control the SSE in FM/2D-TMD/HM heterostructures such as YIG/ML V-$WS_2$/Pt and YIG/ML V-$WSe_2$/Pt. This approach will pave the way for the development of gate-tunable SSE devices that operate at ambient temperature. The new concept of a gate-controlled SEE device is illustrated in Fig. 12d. Since the magnetic properties (magnetization, magnetic anisotropy) of the magnetic TMD monolayers can be modulated by both light and electric field, these two external stimuli can also be harnessed to achieve highly efficient operation of 2D vdW TMD-based spincaloritronic devices. When exploring this new research direction, it is imperative to address some open, important questions regarding the nature of the 2D-TMD magnetism (decoupling defect- and dopant-induced FM contributions to the net magnetization, coercivity, and magnetic anisotropy) and its impacts on spin transport and spin-thermo-transport. It has been shown that the presence of a 2D material like graphene on YIG strongly suppresses a thermally driven spin current via the SSE or spin pumping [230,231], while the presence of a 2D TMD such as the $WSe_2$ monolayer enhances the thermally driven spin current via the SSE [222,223]. The different origins of these two phenomena are still unclear and under debate. The presence of the graphene layer has been reported to reduce the magnetic damping coefficient of the FM layer (YIG) [230], while little is known about the effect of the 2D-TMD layer on the magnetic damping coefficient of the FM layer. It has been theoretically shown that the magnetic damping of the FM layer is inversely proportional to the SSE voltage. So, an emerging question arises: *How does the inclusion of a 2D TMD layer alter the magnetic damping and hence the spin-to-charge conversion efficiency of the FM layer in a FM/2D-TMD/HM based device?* To address this important question, combined ferromagnetic resonance (FMR), FMR-spin pumping, and SSE experiments should be systematically conducted on such FM/2D-TMD/HM heterostructures.

Apart from that, 2D vdW intrinsic ferromagnets such as monolayers of $CrI_3$ [232] and $CrPbTe_3$ [233], as well as the vdW heterostructure $CrI_3/NiCl_2$ [234], have been predicted to show large SSEs. For example, an extremely large SSE value of 1450 $\mu V\ K^{-1}$ can be achieved for monolayer $CrI_3$ and an SSE value of 1320 $\mu V\ K^{-1}$ for monolayer $CrPbTe_3$ at low temperatures (< 100 K). Recently, a theoretical study has shown the largest spin Seebeck coefficient achieved for a $MnCl_3$ monolayer (~1600 $\mu$V/K) in the 50 to 120 K temperature range, which is much greater than that of $CrI_3$ and $CrGeTe_3$ monolayers [235]. Of course, these predictions must be verified experimentally. As discussed earlier, the magnetic properties of these 2D vdW magnets are electrically manipulable, making them desirable for the development of gate-tunable SSE devices as well. However, these devices are restricted to operate at low temperatures (< 100 K) due to low magnetic ordering temperatures in these 2D vdW magnets [232-234]. On a foundational level, on the other hand, exploring SSE and related effects in these 2D vdW magnets advances the understanding of spin-dependent transport and thermo-transport phenomena down to the 2D limit.

It should be noted here that mechanically exfoliated $CrI_3$ [232] and $CrPbTe_3$ [233] monolayers are extremely sensitive to air and their magnetic properties may become degraded or lost during device fabrication, which represents one of the most challenging technical issues in the manufacturing of 2D vdW spintronic and spin-caloritronic devices.

In addition to its outstanding magnetic properties (high saturation magnetization, Curie temperature, and perpendicular magnetic anisotropy), 2D $Fe_3GeTe_2$ exhibits magnetic skyrmions that can be potentially manipulated by electric and optic means [71,72,75,79,80,135,136,236]. Therefore, this 2D material is a very interesting candidate for exploring magnetic skyrmion driven spin-thermo-transport phenomena via the SSE and ANE. It is important to address an emerging question: *How can a thermally driven spin current be manipulated by electric and/or optic control of magnetic skyrmions in a 2D vdW ferromagnet like* $Fe_3GeTe_2$*?* Since mechanically exfoliated layers of $Fe_3GeTe_2$ crystals are very sensitive to air, it is essential to protect the material by capping it with other materials like hBN or Pt. However, the presence of this capping layer might alter the magnetic skyrmion properties of $Fe_3GeTe_2$ and complicate the overall integration of the device. To overcome this, recent progress in MBE growth of $Fe_3GeTe_2$ films has been made [237-239], which provides some distinct advantages such as large-area $Fe_3GeTe_2$ films with tunable Curie temperature (by varying Fe content or film thickness). However, it remains challenging to control desirable chemical compositions, thickness uniformity, and extrinsic effects of substrates used (lattice mismatch), as well as prohibit the formation of undesired phases (due to atomic interdiffusion) during $Fe_3GeTe_2$ film growth. It appears that the presence of a seed layer like graphene is crucial to grow high quality single-phase $Fe_3GeTe_2$ thin films [239]. We outline the exciting opportunities and technical challenges in the field of spin-caloritronics based on 2D vdW magnets and heterostructures (Scheme 2).

**Scheme 2.** Opportunities and challenges in the field of 2D vdW spin-caloritronics.

| Spin-Caloritronics | |
|---|---|
| **Opportunities** | **Challenges** |
| • Electrically tunable spin-thermo-transport (SSE, ANE, etc.) phenomena in 2D-FM/HM and FM/2D-TMD/HM structures (e.g., YIG/ML V-doped $WSe_2$/Pt)<br>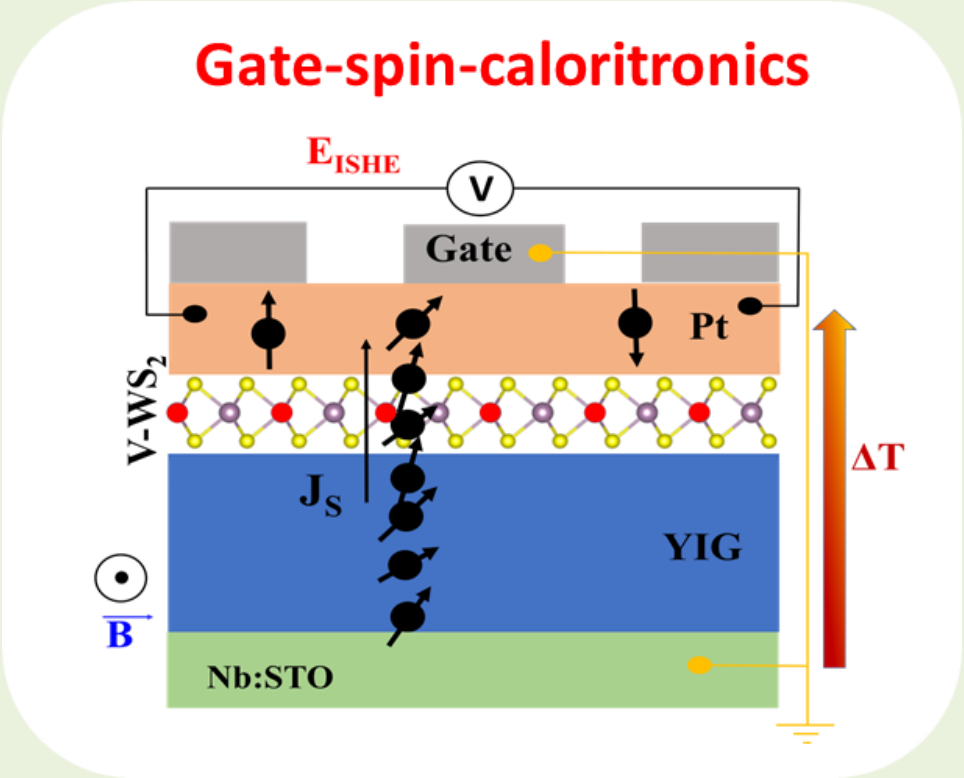<br><br>• Exploring gate-tunable exchange bias or exchange anisotropy in 2D vdW FM/AFM heterostructures (e.g., $Fe_3GeTe_2/FePS_3$) for gate-tunable spincaloritronic devices<br>• Optically tunable magnetism and spin-thermo-transport (SSE, ANE, etc.) in FM/2D-TMD/HM structures (e.g., YIG/ML V-doped $WSe_2$/Pt) for opto-spin-caloritronics | • Weak ferromagnetism<br>• Difficulty controlling magnetic dopants/defects in TMD monolayers resulting in non-reproducible magnetization and hence SSE voltage values.<br>• Air instability (magnetic degradation causes the SSE voltage to reduce)<br>• Intrinsic 2D vdW ferromagnets (e.g., $CrI_3$, $CrPbTe_3$) magnetically order at low temperatures well below 300 K, depending on stacking quality and thickness, air instability (magnetic degradation), and difficulty of device integration<br>• Need to understand spin-charge-phonon coupling mechanisms that govern spin transport across 2D vdW magnetic interfaces<br>• Large discrepancy between the theoretical and experimental values of SSE voltage for the reported 2D vdW magnets and heterostructures |

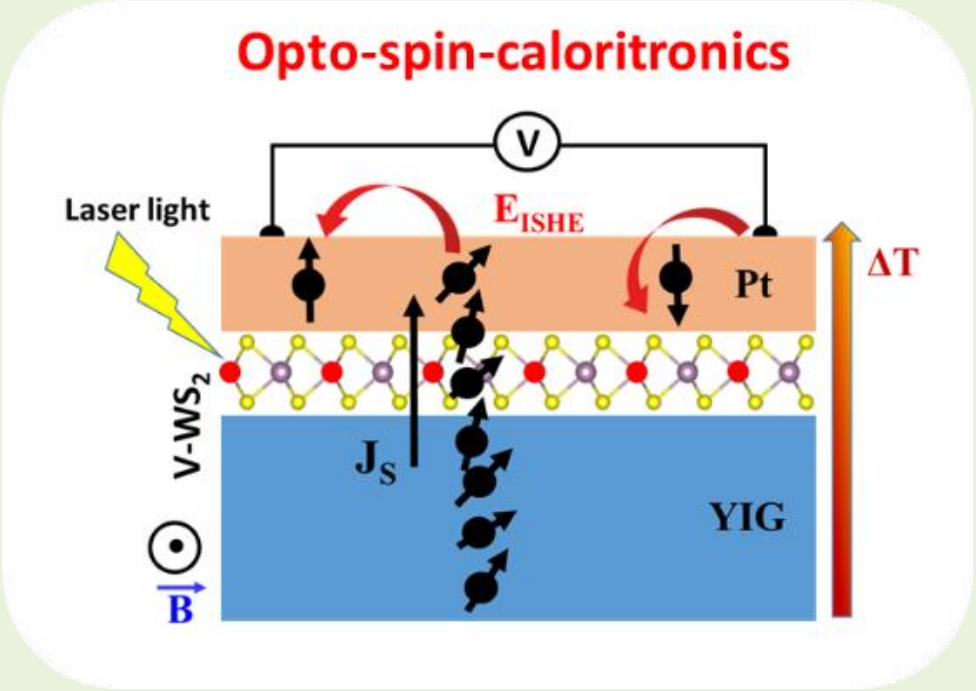

- Ultrafast magnetism and ultrafast spin-thermo-transport for ultrafast opto-spin-caloritronic device applications

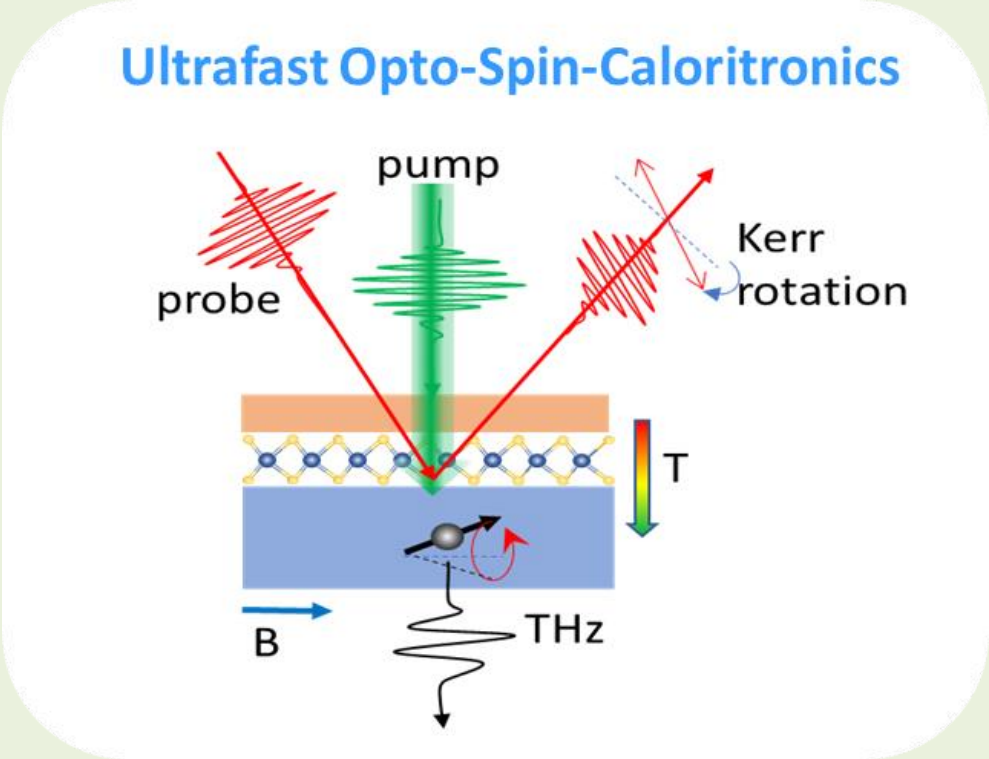

- Electrically and/or optically controllable magnetic skyrmions drive spin current across 2D vdW interfaces (e.g., $Fe_3GeTe_2/WTe_2/Pt$)
- Ferroelectric control of skyrmion states across vdW magnetic/vdW ferroelectric interfaces (e.g., $Fe_3GeTe_2/\alpha\text{-}In_2Se_3$) for spin transport and spin-thermo-transport.

- Need to select vdW materials that order magnetically at temperatures ($T_C$, $T_N$) to achieve EB effects around room temperature
- Difficulty synthesizing and stacking vdW magnetic materials in heterostructures
- Surface and interface oxidization (magnetic degradation)
- Undesired effects of intercalation, interdiffusion at vdW magnet/metal interfaces, twisting, moiré patterns, etc.
- MOKE signals depend on quality of surfaces and interfaces of materials used; nonlocal heating effects on spin dynamics and transport
- In the heterostructure forms, it is challenging to image magnetic skyrmions of 2D magnets subject to external stimuli (gate, light, strain)
- No theoretical models are available to complement experimental findings such as electrically tunable magnetic skyrmions
- Complex device structures; short lifetime of the devices.

### *3.3. Valleytronics*

Valleytronics explores the valley degree of freedom of a 2D-TMD semiconductor and the optic, electric, and magnetic manipulation of its valley polarization [137-147]. While it is rather difficult to polarize valley states using optical approaches, an external magnetic field applied to a 2D-TMD semiconductor can split the valleys due to the Zeeman effect [142,143,178]. To achieve large valley splitting, however, the latter approach requires large magnetic fields (> 10 T), which are usually not accessible in research laboratories [178]. As discussed above, the valley splitting of a semiconducting TMD monolayer (e.g., $WSe_2$, $MoSe_2$) can be enhanced by interfacing it with a 2D vdW magnet (e.g., EuS, $CrI_3$, $Cr_2Ge_2Te_6$, $Fe_3GeTe_2$, $Fe_5GeTe_2$) and the valley polarization of the TMD monolayer can be electronically and/or magnetically controlled through the electric gating of the 2D vdW magnet or its gate-tunable interfacial/exchange magnetic anisotropy [144-147,159,164,165,178,240].

For instance, using $Fe_3GeTe_2$ (FGT) as a ferromagnetic tunnelling contact in the FGT/hBN/$WSe_2$/hBN/graphene heterojunction (Fig. 13a), electroluminescence measurements performed by Li *et al.* [147] revealed the possibility of using bias voltages to inject spin-polarized holes into the monolayer $WSe_2$ (Fig. 13b), causing a population imbalance between the ±K valleys in this 2D-TMD (Figs. 13c,d). The emission of valley excitons occurs upon the application of a magnetic field to the heterostructure under negative (Fig. 13e) and positive (Fig. 13f) bias voltages. The experimental results are complemented by DFT calculations. The authors also observed that under an applied external magnetic field, the helicity of electroluminescence flipped its sign and exhibited a hysteresis loop, the characteristic of which resembled what was obtained from reflective magnetic circular dichroism measurements on the same $Fe_3GeTe_2$ sample [147]. This study opens a new direction for the electric control of valley polarization in 2D-TMDs using vdW ferromagnets like $Fe_3GeTe_2$.

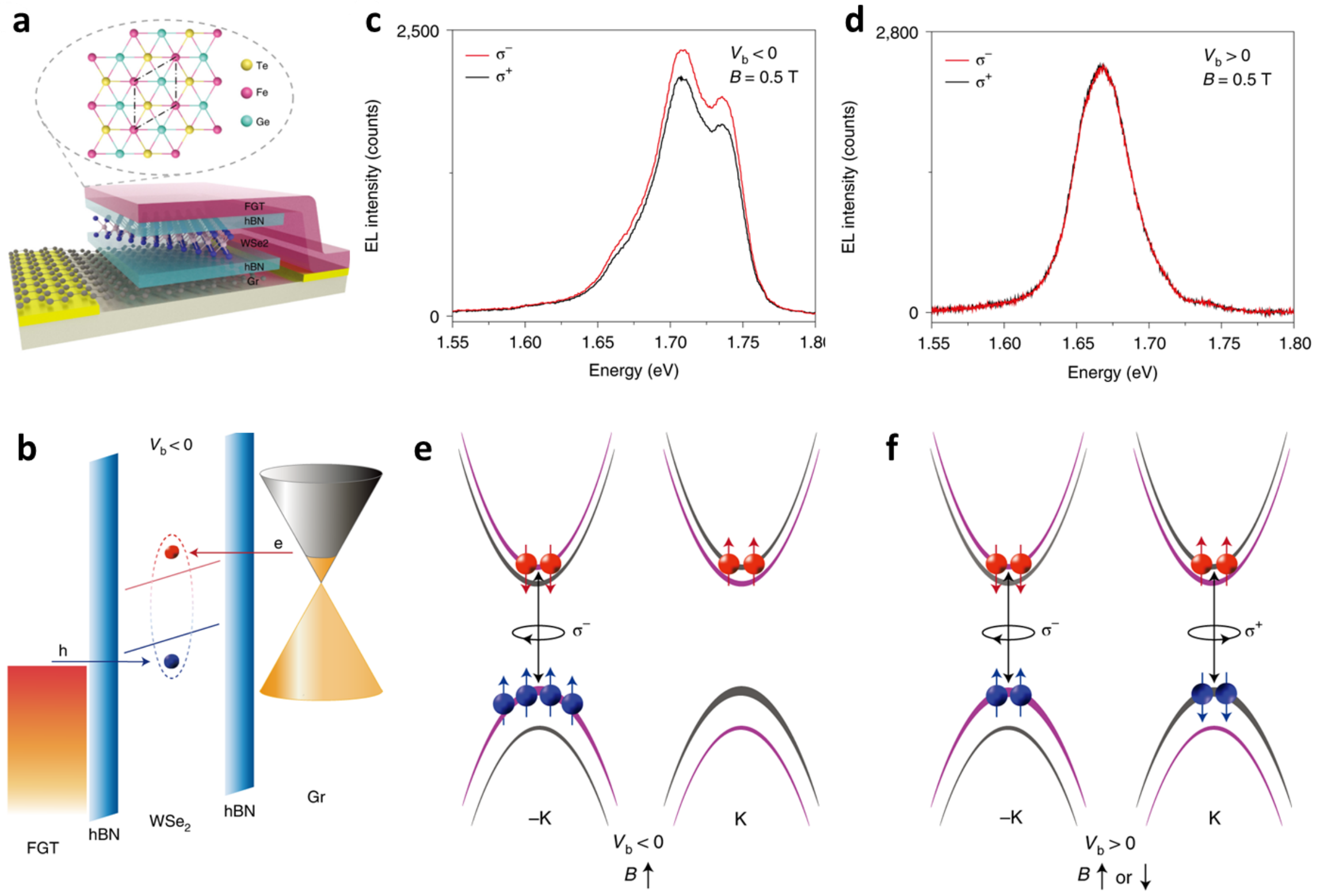

**Figure 13.** (**a**) Schematic of the top view of a vertically stacked FGT/hBN/$WSe_2$/hBN/graphene heterostructure. (**b**) Schematic of the electronic structure of the heterostructure under a negative voltage, $V_b$. Polarization-resolved EL spectra for $\sigma^-$-polarized and $\sigma^+$-polarized detection under **(c)** negative and **(d)** positive voltages (2.4 V) under an applied magnetic field of 0.5 T. Schematic diagrams show the emissions of valley excitons while applying a magnetic field to the heterostructure under (**e**) negative and (**f**) positive voltages. Reproduced with permission from Ref. [147].

As another promising approach, "magnetic doping" has been reported to also enhance the valley splitting in a pristine TMD monolayer (e.g., $WSe_2$, $WS_2$, $MoS_2$), as demonstrated by Li *et al.* for Fe-doped $MoS_2$ monolayers [241] and by Sahoo *et al.* for V-doped $MoS_2$ monolayers [242]. Relative to the pristine $MoS_2$ monolayer, an enhanced valley splitting at 300 K ($g_{eff}$ = -6.4) was observed in the Fe-doped $MoS_2$ monolayer [241]. Similarly, V-doping was found to enhance the valley splitting by 42% in the V-doped $MoS_2$ monolayer as compared to pristine $MoS_2$ (< 12%) [242]. The $g_{eff}$ factor was also tuned to -20.7 by increasing the Fe concentration, which has been attributed to the enhanced Heisenberg exchange interaction of Fe magnetic moments with $MoS_2$ through *d*-orbital hybridization [241]. Since the magnetically-doped TMD monolayers (e.g., V-$WS_2$ and V-$WSe_2$) exhibit strong magnetic responses to both electric field and lasers [52,54,55], the valleytronic properties of these 2D materials can be manipulated by these external stimuli. On the other hand, when placing a magnetically-doped TMD monolayer (V-$WS_2$ or V-$WSe_2$) on top of an intrinsic 2D vdW magnet like $Fe_3GeTe_2$ or $Fe_5GeTe_2$, the MPE induced by this 2D magnet on the electric and magnetic properties of the TMD monolayer may be significant, enabling valley polarization of the TMD monolayer with an appropriately enhanced valley splitting rate. Future research should be conducted to examine this hypothesis.

As noted above, the magnetic properties of TMD monolayers have contributions from defect- and dopant-induced magnetic moments and their couplings [67,123]. So, it would be interesting to investigate effects of transition metal or chalcogen vacancies and magnetic dopant concentrations on the valleytronic properties of the free-standing TMD monolayers, as well as those placed on magnetic substrates in heterostructures. Recently, Liu *et al.* showed that valley splitting can be tuned by varying the ratio of *T* and *H* phases that coexist in a $WSe_2$ monolayer [243]. These studies may open a new approach for defect (atomic vacancy defects) and phase engineering for 2D TMD-based valleytronic device applications.

From a materials' engineering perspective, it is anticipated that heterostructures composed of vdW magnets (e.g., $CrI_3$, $Fe_3GeTe_2$, $Fe_5GeTe_2$) and TMD monolayers (e.g., $WSe_2$, V-$WSe_2$) would provide more atomically flat interfaces as compared to those consisting of non-vdW magnets (e.g., NiFe, YIG, $CoFe_2O_4$) and TMD monolayers (e.g., $WSe_2$, V-$WSe_2$). The strong interfacial couplings between the layers in the former are therefore expected to yield enhanced valleytronic properties through the electrically tunable interfacial charge transfer and/or interfacial magnetic anisotropy mechanisms. In case of the non-vdW magnet/TMD monolayer heterostructures (e.g., YIG/$WSe_2$), lattice mismatch appears to be significant, resulting in the strained TMD monolayer [222,223]. Recent theoretical studies have revealed a significant effect of strain on the magnetic and valleytronic properties of the TMD monolayers [244-248]. An important question arises: *How does strain from a magnetic substrate alter the polarization of valley states in a TMD monolayer?* In case of the vdW magnet/TMD monolayer heterostructures (e.g., $CrI_3$/$WSe_2$, $Fe_3GeTe_2$/$WSe_2$), a "twisting" effect caused stacking two vdW layers could considerably modify their interfacial electronic and magnetic structure and hence their valleytronic functionalities [153,154]. It appears that theoretical and experimental studies have yielded largely different values of the exchange field induced by a magnetic substrate on a TMD monolayer [142,143,178]. This points to the importance of the interface quality (atomically flat interfaces are often considered "perfect" in theoretical calculations but in reality are imperfect in grown samples or in stacked layers) and the difference between the vdW/vdW and non-vdW/vdW interfaces. This discrepancy calls for further studies.

In a similar perspective on "*2D vdW ferroelectric control of 2D vdW magnetism*," it has been theoretically suggested that "2D vdW valleytronics" can be controlled by "2D vdW ferroelectricity" [188,249-252]. By stacking a 2D-TMD (e.g., 2H-$VSe_2$ monolayer) with a 2D vdW ferroelectric material (e.g., $Sc_2CO_2$ monolayer) to form a 2H-$VSe_2$/$Sc_2CO_2$ heterostructure, Lei *et al.* demonstrated gate-tunable valleytronics in this heterostructure [188]. This represents a new approach that may fulfill the increasing requirements of modern valleytronic devices for miniaturization, low-dissipation, and non-volatility. While most of the works reported to date have been based on theoretical calculations, experimental studies are needed to validate these predictions [249-252]. The lack of experimental support is partially due to difficulty growing 2D vdW ferroelectric (FE) materials and stacking them with 2D vdW ferromagnets in heterostructures with atomically controlled interfaces [250,251]. As "twisting" may occur while stacking 2D TMD/2D vdW FE layers, it is essential to investigate twisting effects on ferroelectricity-valley coupling and hence the valleytronic property of the heterostructure. Guan *et al.* provided a list of promising 2D vdW ferroelectric materials [250]. We outline several exciting opportunities and challenges in the field of 2D vdW valleytronics (Scheme 3).

**Scheme 3.** Opportunities and challenges in the field of 2D vdW valleytronics.

| **Valleytronics** | |
|---|---|
| **Opportunities** | **Challenges** |
| • Electrically and optically tunable valleytronics based on 2D room temperature ferromagnetic TMD semiconductors (e.g., V-$WSe_2$, V-$MoSe_2$) through magnetic doping<br>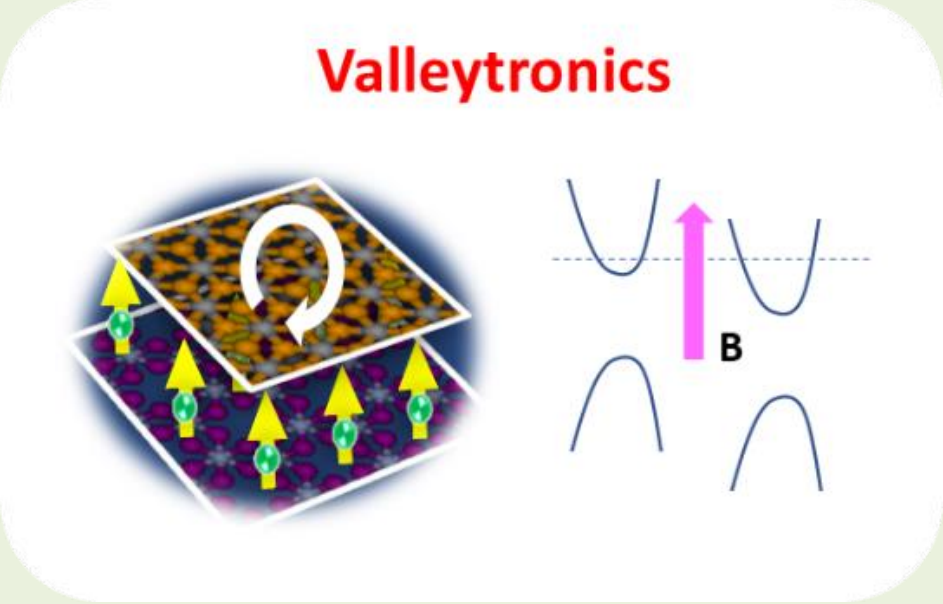<br><br>• Electrically tunable valleytronics in 2D vdW magnet/2D-TMD heterostructures (e.g., $Fe_3GeTe_2$/V-$WSe_2$, $Fe_5GeTe_2$/V-$WSe_2$) through magnetic proximity and/or charge transfer<br>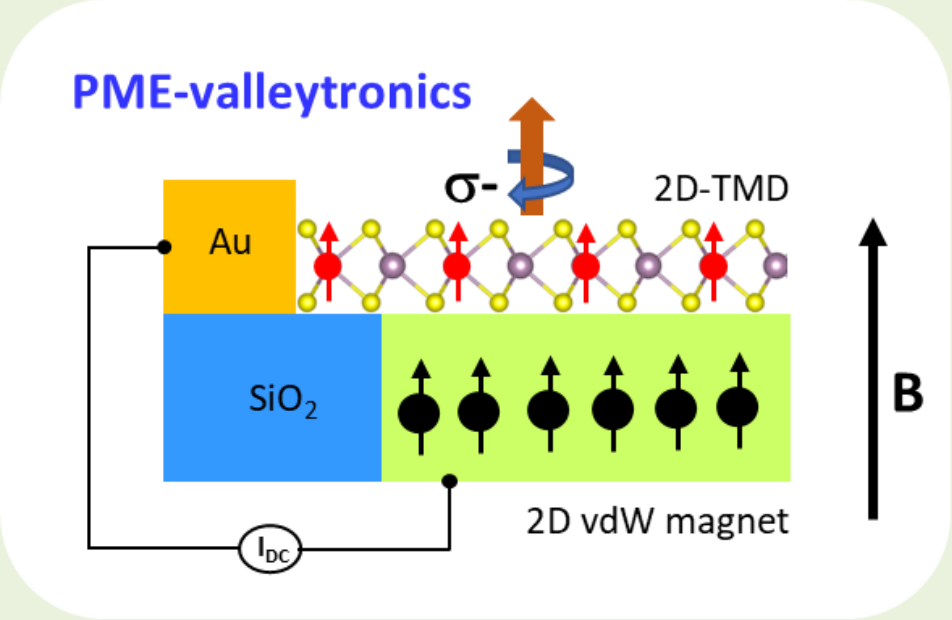<br><br>• Electric control of magnetic skyrmions that | • Difficulty controlling magnetic dopants/defects in 2D-TMDs to achieve reproducible magnetic moment and valley polarization values<br>• Decoupling vacancy defect- and dopant-induced magnetic contributions to the net magnetization, spin-valley coupling, and valley polarization<br>• Air instability (degradation of magnetic and valleytronic properties of TMD monolayers once exposed to air)<br>• Complex origins of the valley polarization of a magnetically-doped TMD monolayer adjacent to a magnetic substrate arise from interrelated magnetic dopants, magnetic proximity, and charge transfer contributions<br>• Need appropriate theoretical models to complement experimental findings on valley-splitting enhancements |
| drive 2D vdW spin-valleytronics (e.g., $Fe_3GeTe_2/WTe_2$)<br>• Ferroelectric control of 2D vdW valleytronics (e.g., $Cr_2Ge_2Te_6/\alpha\text{-}In_2Se_3$, $MnPS_3/CuInP_2S_6$)<br>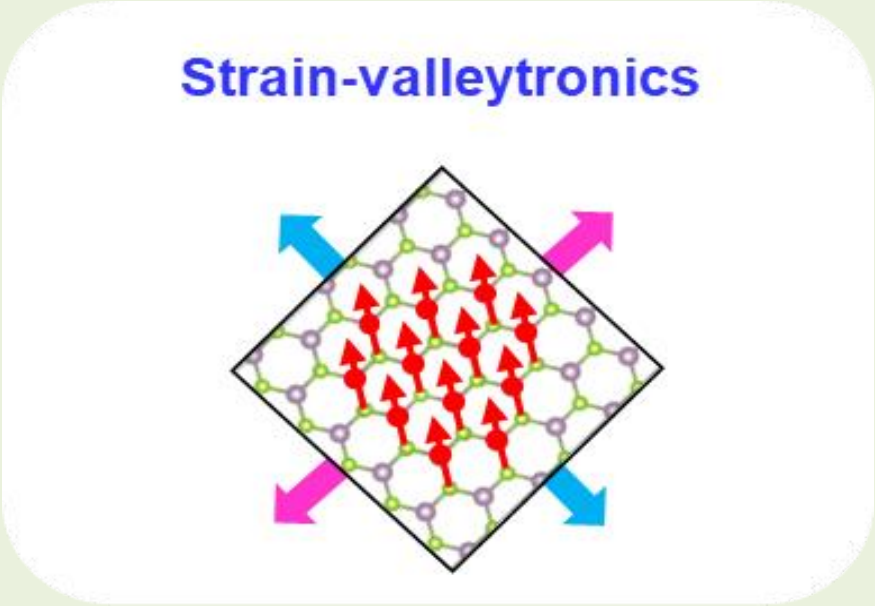<br><br>• Strain control of 2D vdW valleytronics in magnetic 2D-TMDs and heterostructures | • Dependence on stacking, twisting, and thickness; difficulty with device integration and fabrication<br>• Difficulty assembling TMD monolayers and 2D vdW ferroelectric materials in 2D TMD/2D FE heterostructures<br>• Interatomic diffusion at interfaces; surface/interface oxidization (reduced spin-valley or ferroelectricity-valley or strain-valley couplings) |

## 4. Conclusions and Outlook

The electric (current, electric field) control of magnetism in 2D vdW magnets and their heterostructures has provided new and exciting opportunities for next-generation device applications such as spin transistors, magnetic and magnetoelectric sensors, and magnetic memory storage technology. Relative to conventional non-vdW magnetic systems, the control of magnetization or the switching of magnetization becomes easier within 2D vdW magnets, especially when spintronic devices urgently demand further downsizing. For magnetic information storage, sufficiently high magnetic anisotropy is required to stabilize information. On the other hand, magnetic anisotropy should be small for writing information with low energy consumption. Such contradictory requirements represent a grand challenge to which a solution using non-vdW magnetic materials is hardly reached. Since the magnetic anisotropy of 2D vdW magnets is largely tunable by small gate voltages, the electric manipulation of magnetic anisotropy can enable a high working efficiency of spintronic devices. To this end, the present review has provided an in-depth analysis of strategies for electrically controlling the magnetization, magnetic ordering, coercivity, and magnetic anisotropy in 2D vdW magnets and their heterostructures. It further assesses the gate-tunable magnetic properties of emerging 2D vdW intrinsic magnets, including $CrI_3$, $Cr_2Ge_2Te_6$, $Fe_3GeTe_2$, and $Fe_5GeTe_2$ [81,87,88,110-113,118-120,132]. While 2D vdW magnetic semiconductors $CrI_3$ and $Cr_2Ge_2Te_6$ show outstanding electrically controllable magnetic properties below 100 K, the ionic gating of 2D vdW magnetic metal $Fe_3GeTe_2$ has shifted the Cutie temperature (from ~230 K for bulk $Fe_3GeTe_2$) to room temperature and enhanced the saturation magnetization, making it a promising candidate for applications in spintronics and valleytronics near room temperature [118-120]. However, $Fe_3GeTe_2$ or $Fe_5GeTe_2$ suffers from poor air stability, which causes its magnetic properties to degrade rapidly when exposed to air [93]. Careful attention must thus be paid when preparing these 2D materials for spintronic devices. The metallic nature of $Fe_3GeTe_2$ or $Fe_5GeTe_2$ also limits access to a rich variety of optic, opto-electronic, and magneto-optic phenomena open only to semiconductors. Thus, magnetically-doped TMD monolayers emerge as a novel class of 2D diluted magnetic semiconductors. Their electrically- and optically-tunable magnetic properties make them a promising candidate for use in 2D vdW spintronic devices enabling room temperature operation [50-55]. These 2D vdW magnets can also be used as a novel 2D spin filter to boost spin-to-charge conversion efficiency via the SSE in FM/2D-TMD/HM systems [222,223]. For applications with this type of spincaloritronic device, SSE voltage can be controlled by light irradiation, electric gating, or a combination of both, which has been experimentally demonstrated to tune 2D TMD magnetism [52,54,55]. A clear understanding of spin-charge-phonon coupling mechanisms in 2D-TMDs and heterostructures will be the key to open the new fields of "Opto-Spin-Caloritronics" and "Gate-Spin-Caloritronics."

As a promising alternative, the ferroelectric polarization of a 2D vdW ferroelectric material can be used to switch the magnetization or control the magnetic anisotropy of a 2D vdW ferromagnet [136,159,188,190,191]. Since this approach has been mostly based on DFT predictions [188,190,191], experiments are needed to validate the hypothesis. Similarly, the ferroelectric polarization of a 2D vdW ferroelectric material can be used to control the creation and annihilation of magnetic skyrmions in 2D vdW magnets to drive spin currents in spintronic and spincaloritronic devices [72]. However, such 2D vdW magnetic skyrmions are rather unstable in air, which may hamper their practical implementation. A recent study has shown that the application of an external magnetic field perpendicularly to an air-stable $Cr_3Te_4$ nanosheet can transform magnetic stripe domains into biskyrmion bubbles whose size can be varied by magnetic fields [253]. Controlling 2D vdW magnetism by 2D vdW ferroelectricity represents a new, exciting approach for electrically tunable 2D vdW magnetism, however, research in this direction is still in its infancy and needs further exploration.

Harnessing spin-valley, ferroelectricity-valley, and strain-valley couplings to split valley states in 2D-TMDs will likely promote and expand valleytronic applications to include information processing and quantum communications. Novel approaches utilizing advantages of magnetic substrates (via magnetic proximity), magnetic doping, electric field, carrier doping, and phase engineering have been proposed for this purpose [137-147]. It appears that an appropriate combination of these approaches will enable the implementation of multifunctional 2D-TMD materials for next-generation valleytronic devices [140,146,254].

While most of the research reviewed here focused on 2D vdW ferromagnets, recent works have shown that 2D vdW antiferromagnets are also promising candidates for the development of gatable spintronic devices [115,255-257]. Chen *et al.* fabricated magnon valve devices using the AFM insulator $MnPS_3$ and showed that the second harmonic thermal magnon signal of the device could be tuned from positive to negative when a dc electric current was applied through a gate electrode [255]. This discovery may find an important application in 2D magnon-based information technology. Cai *et al.* recently showed that $MnBi_2Te_4$ is a remarkable canted-antiferromagnetic (c-AFM) Chern insulator, and its electrically controlled functionalities are desirable for application in topological c-AFM spintronics [256].

From an engineering perspective of 2D vdW materials, it should be noted that when assembling or stacking vdW layers, intercalation, interdiffusion, twisting, and moiré patterns, etc. may arise [114,115,154]. These effects, in addition to magnetic proximity and charge transfer that likely occur in such heterostructures [221], might reduce or enhance the magnetic responses of the materials to external stimuli (electric field, light, strain) [142,213-216]. It is anticipated that the future research of 2D vdW spintronics, spin-caloritronics, and valleytronics will be driven by new discoveries of air-stable 2D vdW magnets with large magnetic moments and high Curie temperatures around room temperature [258-261]. Establishing new theoretical models is also essential to complement these experimental discoveries.

**Acknowledgments**

Research was supported by the U.S. Department of Energy, Office of Basic Energy Sciences, Division of Materials Sciences and Engineering under Award No. DE-FG02-07ER 46438. The author acknowledges Dr. Tatiana Eggers for her proofreading of the manuscript and the useful comments.